\documentclass{article}

\usepackage[final]{corl_2020} 

\usepackage{multicol}
\usepackage{xcolor}

\usepackage{graphicx, amsmath, subcaption, amssymb, comment, bm, amsthm, bbm}
\usepackage{algorithmic}
\usepackage{algorithm}
\usepackage{siunitx}
\usepackage{soul}
\usepackage{subcaption}
\usepackage{wrapfig}

\hypersetup{
    colorlinks=true,
    linkcolor=blue,
    filecolor=magenta,      
    urlcolor=cyan,
}
\usepackage{soul}

\newcommand{\algorithmicinput}{\textbf{Input:}}
\newcommand{\algorithmicoutput}{\textbf{Output:}}

\newcommand{\INPUT}{\item[\algorithmicinput]}
\newcommand{\OUTPUT}{\item[\algorithmicoutput]}

\newcommand{\hpp}{\texttt{HPP}}
\newcommand{\tuple}[1]{\langle #1 \rangle}

\newcommand{\task}{rendezvous task}
\newcommand{\mdp}{\mathcal{M}}
\newcommand{\numagents}{n}
\newcommand{\states}{\mathcal{S}}
\newcommand{\observations}{\mathcal{O}}
\newcommand{\actions}{\mathcal{A}}
\newcommand{\dynamics}{T}
\newcommand{\reward}{\mathcal{R}}
\newcommand{\action}{\bm{a}}
\newcommand{\highRate}{T_h}
\newcommand{\lowRate}{T_l}
\newcommand{\pose}{\bm{p}}

\newcommand{\lidar}{\bm{o}}
\newcommand{\goal}{\bm{g}}
\newcommand{\lv}{v}
\newcommand{\av}{\theta}

\newcommand{\ptwop}{\bm{\pi}}
\newcommand{\hp}{\bm{\Pi}}
\newcommand{\goaldistance}{d}
\newcommand{\alg}{\hpp}
\newcommand{\f}{\bm{f}}
\newcommand{\gauss}[2]{\mathcal{N}(#1 , #2^2)}
\newcommand{\posemu}{\pose_\mu}
\newcommand{\posestd}{\pose_\sigma}

\newcommand{\D}{\mathcal{D}}

\newcommand{\fipose}{\Delta\pose}
\newcommand{\fiobservation}{\Delta\lidar}
\newcommand{\fiobservationmi}{\Delta\currentobservationmi}

\newcommand{\currentpose}{\hat\pose}
\newcommand{\currentobservation}{\hat{\lidar}^{(s)}}
\newcommand{\currentobservationmi}{\tilde{\lidar}^{(s)}}

\title{Model-based Reinforcement Learning \\ for Decentralized Multiagent Rendezvous}

\author{
    Rose E. Wang\thanks{Corresponding author. The research was conducted during Rose's internship at Robotics at Google.}\\
    Google Research, Stanford University  \\
    \texttt{rewang@stanford.edu}\\
    \And
    J. Chase Kew \\
    Robotics at Google\\
    \texttt{jkew@google.com}\\
    \And
    Dennis Lee\\
    Google Research\\
    \texttt{ldennis@google.com}\\
    \And 
    Tsang-Wei Edward Lee\\
    Robotics at Google\\
    \texttt{tsangwei@google.com}\\
    \And 
   Tingnan Zhang\\
   Robotics at Google\\
   \texttt{tingnan@google.com}\\
    \And 
   Brian Ichter\\
   Robotics at Google\\
   \texttt{ichter@google.com}\\
    \And 
   Jie Tan\\
   Robotics at Google\\
   \texttt{jietan@google.com}\\
    \And 
   Aleksandra Faust\\
   Google Research\\
   \texttt{faust@google.com}

}

\begin{document}
\maketitle


\begin{abstract}
    Collaboration requires agents to align their goals on the fly. Underlying the human ability to align goals with other agents is their ability to predict the intentions of others and actively update their own plans. We propose hierarchical predictive planning (\hpp), a model-based reinforcement learning method for decentralized multiagent rendezvous. Starting with pretrained, single-agent point to point navigation policies  and using noisy, high-dimensional sensor inputs like lidar, we first learn via self-supervision motion predictions of all agents on the team. Next, \hpp\ uses the prediction models to propose and evaluate navigation subgoals for completing the rendezvous task without explicit communication among agents.
    We evaluate \hpp\ in a suite of unseen environments, with increasing complexity and numbers of obstacles. We show that \hpp\ outperforms alternative reinforcement learning, path planning, and heuristic-based baselines on challenging, unseen environments.
    Experiments in the real world demonstrate successful transfer of the prediction models from sim to real world without any additional fine-tuning. Altogether, \hpp\ removes the need for a centralized operator in multiagent systems by combining model-based RL and inference methods, enabling agents to dynamically align plans.\footnote{The video is available at: https://youtu.be/-ydXHUtPzWE}
\end{abstract}



\section{Introduction}
Imagine you and your friend plan to meet up, and you find yourselves at two ends of a busy crosswalk. How do you efficiently meet at a common location? There are several possibilities that come to mind and might be efficient ways of accomplishing this: Either you could wait for your friend to join you on your side (or vice versa) or you both can attempt to meet in the middle. But what about the people in your way? You attempt to weave around the group by turning left, but you see your friend heading in the opposite direction. Without communication, you understand that both of you imagined different locations and so you replan your route to meet them more quickly.

\begin{figure}[th]
    \centering
    \begin{subfigure}{.45\textwidth}
        \includegraphics[height=3.5cm]{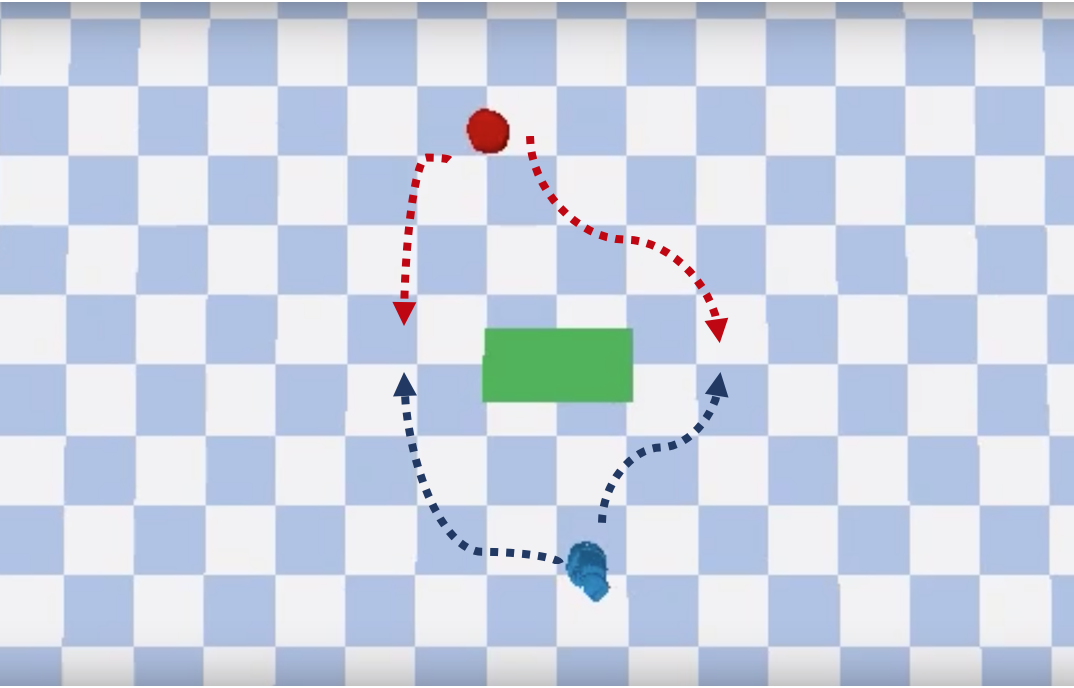}
        \caption{\label{fig:example_setting}}
    \end{subfigure}
    \begin{subfigure}{.48\textwidth}
        \centering
        \includegraphics[height=3.5cm]{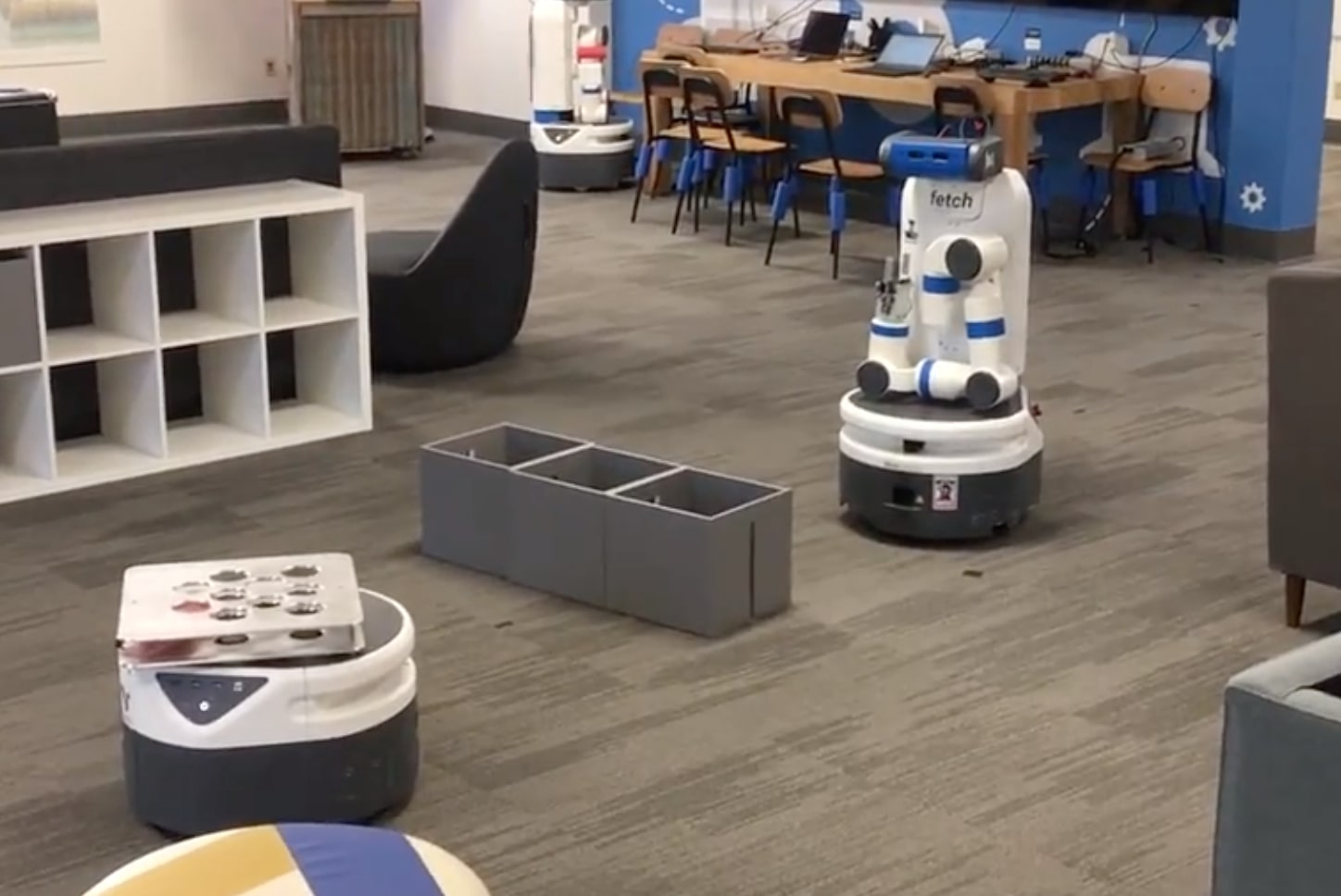}
        \caption{\label{fig:real_world}}
    \end{subfigure}
    \caption{\small \textbf{(a)} Top down view of two independently controlled robots ({\color{red}top}, {\color{blue}bottom}) separated by obstacles ({\color{green}center}) must meet each other. How should they move in order to meet? Example trajectories are illustrated in dashed arrows with the robot's corresponding colors. \textbf{(b)} Real-robot experiments performing \task\ with zero-shot sim2real transfer.}
\vspace{-1.5\baselineskip}
\end{figure}

The above example is the focus of this paper and an instance of a decentralized \textit{\task} \cite{durrant2006simultaneous,faust-icra-15}, where agents must align their goals \cite{wang2020too,tumer2010aligning,tumer2002learning} without explicit communication. The \task\ plays an important role in real world multiagent and human-robot settings for e.g. performing object handovers \cite{strabala2013toward, huang2015adaptive}. While other works address reliable sensor-informed goal navigation \cite{autorl, unstuck-dinesh}, the problem of aligning goals is nontrivial across independent agents with different beliefs without a centralized planner \cite{brunet2008consensus} (which isn't always readily available in the real world). The selection of a good rendezvous point depends on the obstacles in the environment, and on the policies and dynamics of the agents. Take for example Fig. \ref{fig:example_setting} where two agents must meet. Without communication, each agent needs to interpret a stream of high-dimensional, noisy sensor inputs to align goals with other agents, while keeping in mind the navigational obstacles and handling miscoordinations, such as agents heading towards goals in opposite directions. Thus, agents must adaptively coordinate \cite{stone2010ad} to align goals. The obstacle between the agents might prompt a miscoordination, such as the red agent moving west and the blue agent east. Resolving this miscoordination depends on agents' ability to model others' motions and to adapt to diverging intentions using the limited information. 

Assuming a decentralized, real-world multiagent system, there are key questions in modelling, predicting, and accounting for other agents' behavior. These questions are three-fold: How should agents coordinate using high-dimensional and imperfect sensor inputs? How should agents cooperate with others under partial observability and uncertainty? How should agents resolve miscoordinations?  
Our contribution is a holistic approach to address these challenges. We present a decentralized, learning-based system to solve the \task. 
Akin to the standard navigation pipeline, our learning-based system consists of three modules: control, prediction, and planning. A robot's control module is equipped with a pre-trained, imperfect navigation policy that can navigate to a given location in the obstacle-laden environment. Using only an agent's own observations and assumed goal, the prediction module learns via self-supervision to predict the motion of agents. The planning module, hierarchical predictive planning (\hpp), is a hierarchical model-based reinforcement learning (RL) method. It selects and updates its beliefs over  rendezvous points: The prediction module evaluates the points, and the evaluations are used to update an agent's beliefs over potential rendezvous points. The planning module outputs a rendezvous point to the control policy for execution. While the hierarchical planning and control setup are not unusual \cite{ayzaan, prm-rl}, our work closes the loop between the control and planning for decentralized multiagent systems through the motion predictors. We believe this work is of interest to the larger a) multiagent community as a real-world example of a decentralized, cooperative task using noisy sensors and imperfect controllers, b) motion planning community as an example of a learning-based planning system that closes the loop between the planner and controller, and c) RL community as an example of model-based RL as feedback in a hierarchical, self-supervised prediction setting. 

\section{Related works}
Self-supervised prediction and model-based RL methods have also helped make planning more sample-efficient \cite{ebert2018visual,hirose2019deep} and more interpretable \cite{polydoros2017survey, buckman2018sampleefficient}. Additionally, prior works have used learned prediction models for goal proposal and selection \cite{rlrrt,somil}. However, these learning methods do not work in the multiagent setting, where reasoning about other agents is key. There are works that learn motion predictive models from a third party point of view \cite{rabinowitz2018machine,jaques2019social,barrett2012learning}, however these works either cannot scale to real-world sensors or cannot be transferred from simulation to reality. There are also several prior works in human motion prediction \cite{chung2011predictive} applied to the collision-avoidance task \cite{everett2019collision,kuderer2012feature}, another example of a non-communicating multiagent navigation problem. However, the \task\ presents a different navigation challenge where methods for collision-avoidance do not apply: it requires agents to coordinate both in time and space to align goals. There has been work to learn coordination end-to-end via the paradigm of centralized training and decentralized execution \cite{lowe2017multiagent, wang2019r}. However, as we show in experiments, this class of methods fails to scale in real-world, more challenging environments. 
Our work is most similar to \cite{wang2020too} which studies decentralized multiagent systems and aligning team goals by dynamically learning about the other agents via inference and model-based planning. However, similar to other decentralized multiagent planning methods \cite{claes2017decentralised, desaraju2011decentralized}, they make assumptions such as having observations of the entire state of the grid world environment (a bird's eye view), access to a perfect model of agents and object interactions (i.e. their model of the world is not learned) and ground truth knowledge of which objects cause collision. In order to bring our robots closer to the  real world, our work does not make these assumptions, making the \task\ and aligning goals across agents more challenging. 

\section{Problem definition and preliminaries \label{section:background}}
We define the problem space as a decentralized partially observable Markov Decision Process (Dec-POMDP) \cite{oliehoek2012decentralized}. The Dec-POMDP, $\mdp,$ is the tuple $\tuple{\numagents, \states,\observations,\actions_{1\ldots n}, \dynamics, \reward, \gamma}.$ $\numagents$ is the number of agents. We adopt the game theory notation $i$ and $-i$, which refer to agent $i$ and all agents except $i$ respectively. $\states$ is a set of states describing the possible configurations for agents and objects in the environment. We assume the true state space to be hidden and we do not explicitly model it. The joint observation space is $\observations = [\observations_1,\ldots,\observations_\numagents],$ where $\observations_i$ is agent $i$'s observations.  $\actions_{1 \ldots n}$ is the joint action space with $\action_i \in \actions_i$ being the set of actions available to agent $i$. The transition function $\dynamics: \states \times \actions_{1 \ldots n} \times \states \rightarrow [0,1]$ is the probability of transitioning from one state to another after each agent takes its respective action $a_{1 \ldots n}$. As with the true system state, we assume the transition function is unknown. The reward $\reward$ maps system state and action to a scalar that represents the collective reward function. Finally, $0 < \gamma < 1$ is the discount factor.

Each robot observes noisy approximations of agent poses, its own sensor observations, and relative position of its goal (selected by its own planner) in polar coordinates, yielding $\observations_i = [\pose_{i}, \pose_{-i}, \lidar, \goal] \in \observations_i.$ Each pose consists of the relative x (meters), relative y (meters) and relative heading (radians) information of the agent. Because our work focuses on dynamically aligning navigation goals of independent agents with prediction and planning, we leave the problem of inferring poses of other agents from local observations for future work. The actions for each robot are the robot's linear and angular velocities $\action_i = [\lv, \av].$



We assume that each agent is capable of single-agent navigation in obstacle-laden environments, using a policy which we refer to as a P2P (point-to-point) policy. These policies are pre-trained in a separate set of environments, and used here as building blocks without additional training.



\begin{figure}[tb]
    \centering
    \begin{subfigure}{.25\textwidth}
        \includegraphics[width=0.9\textwidth]{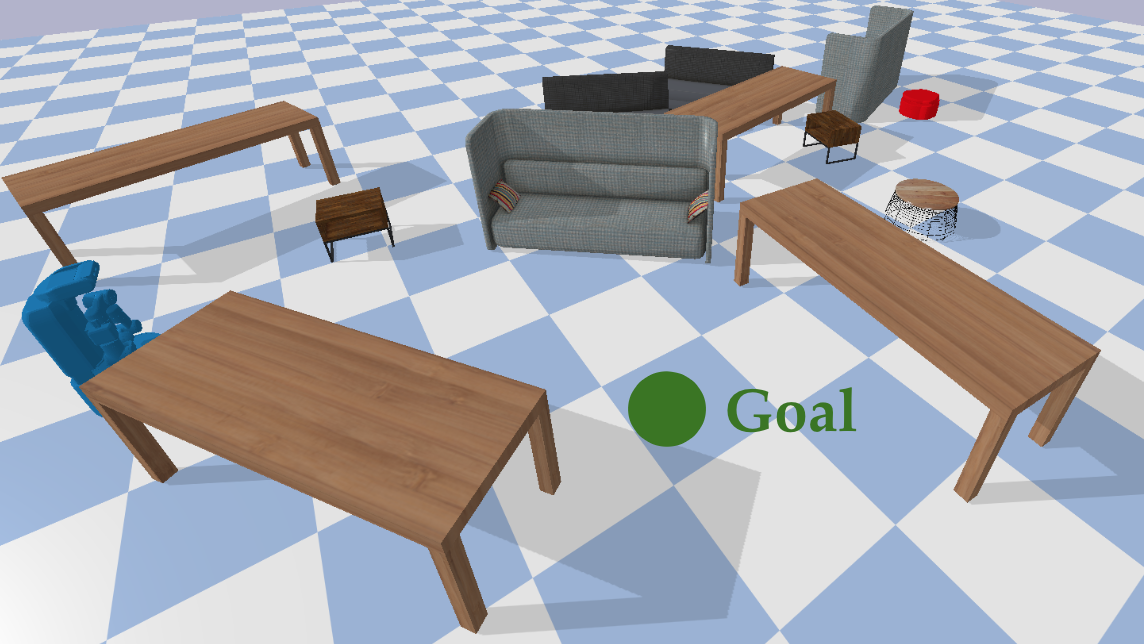}
        \caption{Training environment\label{fig:training_env}}
    \end{subfigure}
    \begin{subfigure}{.25\textwidth}
        \centering
        \begin{minipage}{0.9\textwidth}
            \centering
            \includegraphics[width=\textwidth]{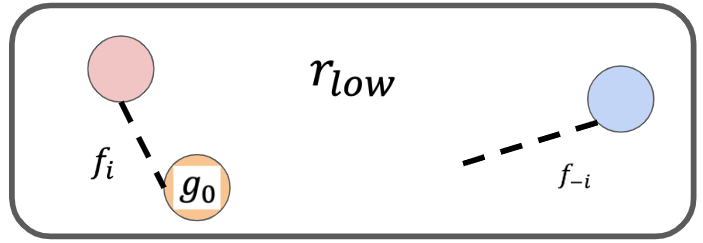}
            \vfill
            \includegraphics[width=\textwidth]{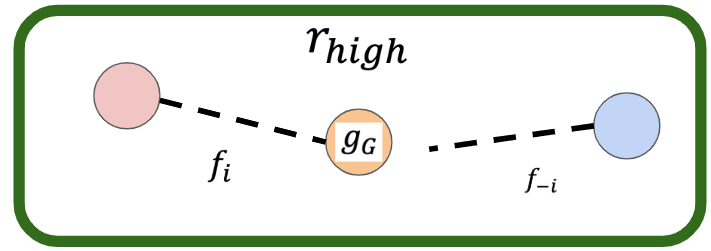}
        \end{minipage}
        \caption{Goal evaluation.\label{fig:goal_evaluation_examples}}
    \end{subfigure}
    \begin{subfigure}{.48\textwidth}
        \includegraphics[width=\textwidth]{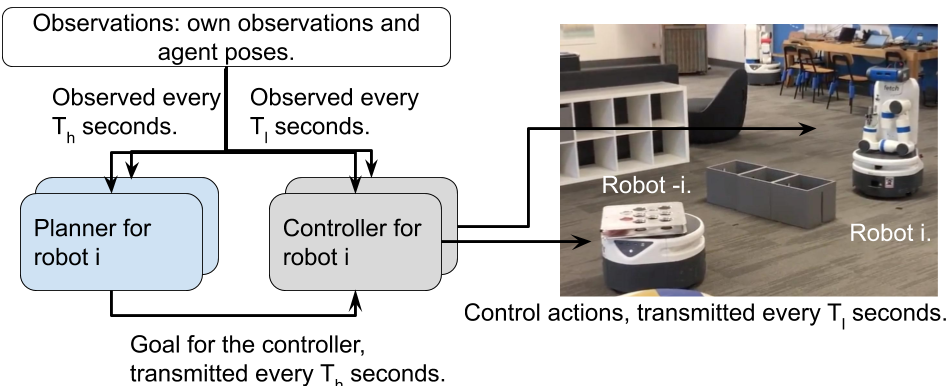}
        \caption{HPP architecture.\label{fig:hpp}}
    \end{subfigure}
    
    \caption{\small \textbf{(a)} Motion prediction model, $\f_i, \f_{-i}$, training environment with randomly filled obstacles. All agents ({\color{blue}{left}}, {\color{red}{upper right}}) are given the same random goal ({\color{green}{center}}) and move with their own P2P policies towards it. \textbf{(b)} Goal ($\goal_0$ and $\goal_G$) evaluation during the high-level policy $\hp$ execution. At the end of a simulated trajectory, the agents (\textcolor{red}{left} and \textcolor{blue}{right}) are either a) far or b) close to each other. $\goal_G$ is a better goal than $\goal_0$ because agents end up closer to each other. \textbf{(c)} \alg\ architecture at run-time. Each agent has a planner and P2P controller running at different frequencies. Each agent receives its own sensor information and agent poses. The planner proposes a goal from the agent's beliefs about how to best align goals among agents. Using the goal, the controller outputs an action for the agent to perform.}
\end{figure}

\section{Methods \label{sec:methods}}
First, the method trains motion prediction models in simulation for the given P2P control policies. This training is done in a set of environments independent from the deployment environments. Section \ref{section:prediction-models} describes the prediction model training. 

Second, after the prediction models are trained, a model-based reinforcement learning planner uses the learned models in the deployment environments to guide the agents towards the rendezvous. 
Each agent runs its own planner and P2P controller (Figure \ref{fig:hpp}). The planner takes the agent's observations and outputs a goal for the controller. The controller takes the planner's goal and agent's sensor observations and outputs a control action to the robot. 
When the planner selects the next goal, it takes into account what it believes the other agents would do if they shared the goal of completing the \task. To perform this reasoning, the planner uses prediction models to predict trajectories. These trajectories are used to evaluate goals against the joint team objective. Section \ref{section:high-level-policy} outlines the algorithm. Together, the model and model-based RL planning enable real-time adjustments, recover on-line from miscoordinations, and lead agents to task completion in unseen environments.



\subsection{Motion prediction model training via self-supervision\label{section:prediction-models}}
The prediction models predict an agent's goal-conditioned motion. 
Similar to \cite{nagabandi2018neural, nagabandi2018learning}, we learn two models, \textit{self-prediction} ($\f_i$) and \textit{other-prediction} ($\f_{-i}$), both of which are trained on observations of an agent following a hidden controller. 
The key difficulty lies in training $\f_{-i}$ without access to all of agent $-i$'s sensor observations, particularly lidar. The model must learn to predict the other agent's likely motion based on its own lidar readings.
The self-prediction model takes in a history length $h$ of the agent's own sensors $\pose_{i}^{t-h:t}, \lidar_{i}^{t-h:t}$ and fixed training goal $\goal$ between time steps $t-h$ to $t$. The self-prediction model outputs $\Delta \pose_{i}^{t+1},\Delta \lidar_i^{t+1}$, i.e. the difference between its next and current pose and sensors with respect to its own policy
\begin{equation}
    (\Delta\pose_i^{t+1},\Delta\lidar_i^{t+1}) = \f_i(\pose_i^{t-h:t}, \lidar_i^{t-h:t}, \goal).
    \label{eq:fi}
\end{equation} 

Similarly, the other-prediction model, which models the motion of agent $-i$, takes in a history length $h$ of agent $-i$'s approximated pose, agent $i$'s own sensors and the fixed training goal. It outputs the change in agent $-i$'s pose and lidar with respect to agent $-i$'s controller,
\begin{equation}
    (\Delta\pose_{-i}^{t+1},\,\Delta\lidar_i^{t+1}) = \f_{-i}(\pose_{-i}^{t-h:t}, \lidar_i^{t-h:t}, \goal).
    \label{eq:fmi}
\end{equation}
Note that the models do not store actions because they learn anticipated poses of agents conditioned on goals. This approach avoids having to know the exact action spaces of robots. 



\paragraph{Data collection \label{sec:data_collection}} Given the P2P controllers $\ptwop_i,\, i=1\ldots \numagents$ and a goal, we collect examples of agents in the environment executing their $\ptwop_i$'s to move to the goal. To provide diverse experiences, we randomize the obstacles in our multiagent simulation environment (see Figure~\ref{fig:training_env} for an example). For each agent, we collect two datasets $\D_i$ (for $\f_{i}$) and $\D_{-i}$ (for $\f_{-i}$) which contain the training information for the prediction models explained previously.
\begin{align*}
\D_i &= [(\pose_i^{t-h:t}, \lidar_i^{t-h:t}, \goal), (\Delta \pose_i^{t+1},\,\Delta \lidar_i^{t+1})] \\
\D_{-i} &= [(\pose_{-i}^{t-h:t}, \lidar_i^{t-h:t}, \goal), (\Delta\pose_{-i}^{t+1},\,\Delta\lidar_{i}^{t+1})].
\end{align*}

\paragraph{Model training} $\D_i$ and $\D_{-i}$ are used to train the \textit{self-prediction} ($\f_i$) and \textit{other-prediction} ($\f_{-i}$) models. Both predictors are approximated with deep neural networks consisting of four fully connected layers, trained with mean squared loss.

\begin{algorithm}[bht]

\begin{algorithmic}[1]
 \INPUT Pretrained prediction models $\f_i, \f_{-i}$; Observations $\observations_i = (\pose_i, \pose_{-i}, \lidar_i, \goal_i)$\\
 \OUTPUT Agent's new goal $\hat\goal_i$\\
 \STATE Initialize goal distribution $\gauss{\posemu}{\posestd}$ using agent poses $\pose_i, \pose_{-i}$
 \FOR{$l < \text{MaxIterations}$ \textbf{if} $\posestd > \epsilon$}
    \STATE Sample $N$ goals $G$ from current distribution \\
    \FORALL{$\goal_j \in G$}
        \STATE Initialize predicted poses $\currentpose_i, \currentpose_{-i}$ \& sensors $\currentobservation_i, \currentobservationmi$ with observation $\observations_i$
        \FOR{k=1\ldots \text{T}}
            \STATE Predict changes in pose $\fipose_i, \fipose_{-i}$ \& sensors $\fiobservation, \fiobservationmi$ with Eq. \eqref{eq:fi} and \eqref{eq:fmi} \\
            \STATE Update $\currentpose_i, \currentpose_{-i}, \currentobservation_i, \currentobservationmi$
        \ENDFOR
        \STATE Compute reward for $\goal_j = \reward(\currentpose_i, \currentpose_{-i})$ using Eq.~\eqref{eq:reward}
    \ENDFOR
    \STATE Select $M$ goals with highest rewards and update $\gauss{\posemu}{\posestd}$
 \ENDFOR
 \RETURN $\posemu$
 \end{algorithmic}
 \caption{\hpp~'s high-level policy, $\Pi_i$.}
 \label{high_level_policy}
\end{algorithm}

\subsection{HPP planner \label{section:high-level-policy}}
The decentralized policy $\hp_i$ uses 1 $\f_i$ and $(\numagents - 1)$ $\f_{-i}$ prediction models to plan and evaluate goals. We use the cross-entropy method (CEM) \cite{de2005tutorial} to convert goal evaluations into belief updates over potential rendezvous points. 
Algorithm~\ref{high_level_policy} describes $\hp_i$ that runs on an agent $i$. The intuition behind $\hp_i$ is that each agent, independently, simulates a fictitious centralized agent that fixes the goal of all agents (Lines 4-8). The goal pre-conditions the motion predicted by $\f_i$ and $\f_{-i}$. Conditioned on a proposed goal, the algorithm predicts the poses of agents for a horizon of $T$ time steps in the future (Lines 6-9). The poses are generated from sequential roll outs of the prediction models $\f_i$ and $\f_{-i}$. Each goal is then evaluated by scoring the anticipated system state using the rendezvous task reward shown in Eq.~\eqref{eq:reward} (Line 10). The reward is the negative difference in agent positions, and is 0 if the task is completed, which happens when agents meet within a predetermined distance, $\goaldistance$, from each other. Specifically, the reward is defined as 
\begin{equation}
     \reward(\pose_{1..n}) = 
    \begin{cases}
        0 & |\pose_{j} - \pose_\mu| < \goaldistance \quad \forall j \in 1..n \\
        \sum_{j, k\neq j } -|\pose_k  - \pose_j| & otherwise
    \end{cases} \text{ where }  \pose_\mu = \frac{1}{\numagents}\sum_{k \in 1..n} \pose_k 
    \label{eq:reward}
\end{equation}
is the center point of the robots' poses (average of their locations), $\goaldistance$ controls the precision of the rendezvous, $\numagents$ is the number of agents. See Appendix \ref{app:reward} for more details.

The planner uses the rewards to update the distribution over goals to favor ones that bring agents closer together (Line 12). 
Figure~\ref{fig:goal_evaluation_examples} illustrates this process: $\hp_i$ evaluates the different goals for each agent by checking the reward it expects to receive after a time horizon given each goal, and it assigns higher reward to the goal that closes the distance between agents.

Finally, to run the \alg\ and complete a coordinated \task\ without coordination, each agent $i$ runs an instance of $\hp_i$. The input to $\hp_i$ is $\numagents$ prediction models: one self-prediction and $\numagents - 1$ other agents prediction models, as each agent might be running a different P2P policy. Every $\highRate$ timesteps, $\hp_i$ observes its sensors, receives poses of all agents, and outputs a recommended goal for agent $i$; in other words, $\highRate$ is the frequency of high-level planner, which controls how frequently the goal is updated).  P2P policy receives the goal and drives the agent to the goal, performing actions every $\lowRate < \highRate$ seconds; in other words, $\lowRate$ is frequency of low-level controller. The process stops when the agents are close to each other or time runs out. 

\section{Results \label{section:experiments}}
To evaluate that presented model we answer the following questions:  1) Does \alg\ leverage the prediction models to align goals? 2) Does \alg\ effectively reduce the rendezvous time? 3) Does \alg\ generalize and effectively handle the rendezvous task in real environments with zero-shot transfer? But first, we describe the experiment setup, baselines and environments.


\textbf{Setup} We use Fetch and Freight mobile robots \cite{wise2016fetch}. The observations are 2D lidar with 222 rays. 
The planning parameters for \alg\ are $T=5$, $\lowRate=1$, $\highRate=10$, which were selected based on the ablation studies. The $\epsilon$ convergence criteria for CEM is 0.001. Appendix~\ref{app:robot_setup} contrain more details.


\textbf{Baselines}
We compare \alg\ to learned, planning, and centralized baselines. MADDPG \cite{lowe2017multiagent} is the learned baseline because it is one of the most popular model-free multiagent RL algorithms (see Appendix~\ref{app:maddpg} for the training details). RRT \cite{lavalle-rrt} is the planning baseline. We combine \texttt{RRT} with CEM (\texttt{RRT+CEM}), which simulates the motion of agents using \texttt{RRT} and selects goals using \texttt{CEM} as described in Section~\ref{section:high-level-policy}. Lastly, the centralized baselines rely on heuristic to fix the goals of agents to a common location -- agents do not perform inference or maintain uncertainty of where to go like \alg\ agents do. Specifically, the centralized baselines set the high level planner's goal to the midpoint of the agents (\texttt{Centralized MP}), the other agent's position (\texttt{Centralized OA}) or a random point in the environment (\texttt{Centralized RP}). For example, \texttt{Centralized MP} and \texttt{RRT+MP} means that no goal inference is performed and agents move towards the midpoint of their positions using our low-level policy or \texttt{RRT} respectively. 

\begin{figure*}[tb]
	\centering
    \vspace{2mm}
	\begin{tabular}{ccc}
	\subfloat[\scriptsize Simple, no obstacle world. ]{\includegraphics[width=0.3\textwidth,keepaspectratio=true]{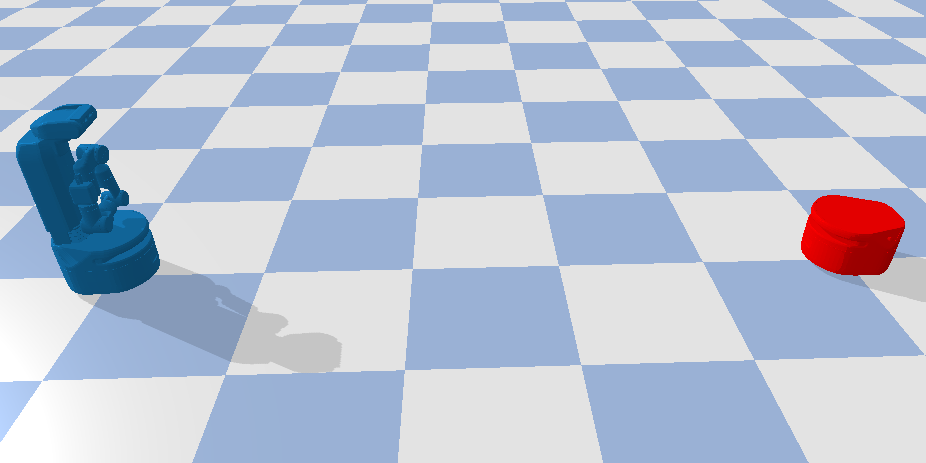}\label{fig:simple_world}} &
	\subfloat[\scriptsize Wall world. ]{\includegraphics[width=0.3\textwidth,keepaspectratio=true]{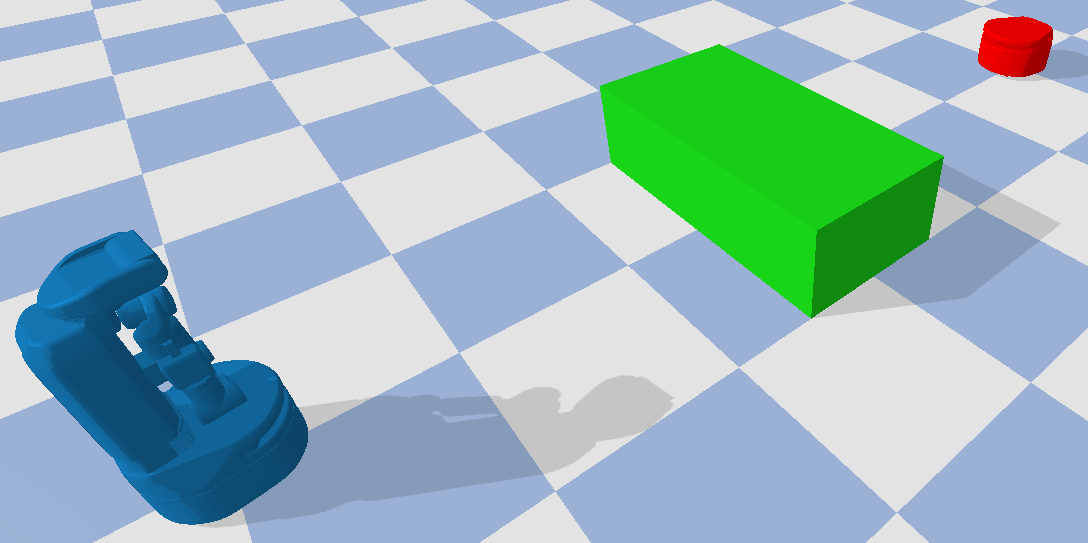}\label{fig:wall_world}} &
	\subfloat[\scriptsize Hard navigation world. ]{\includegraphics[width=0.3\textwidth,keepaspectratio=true]{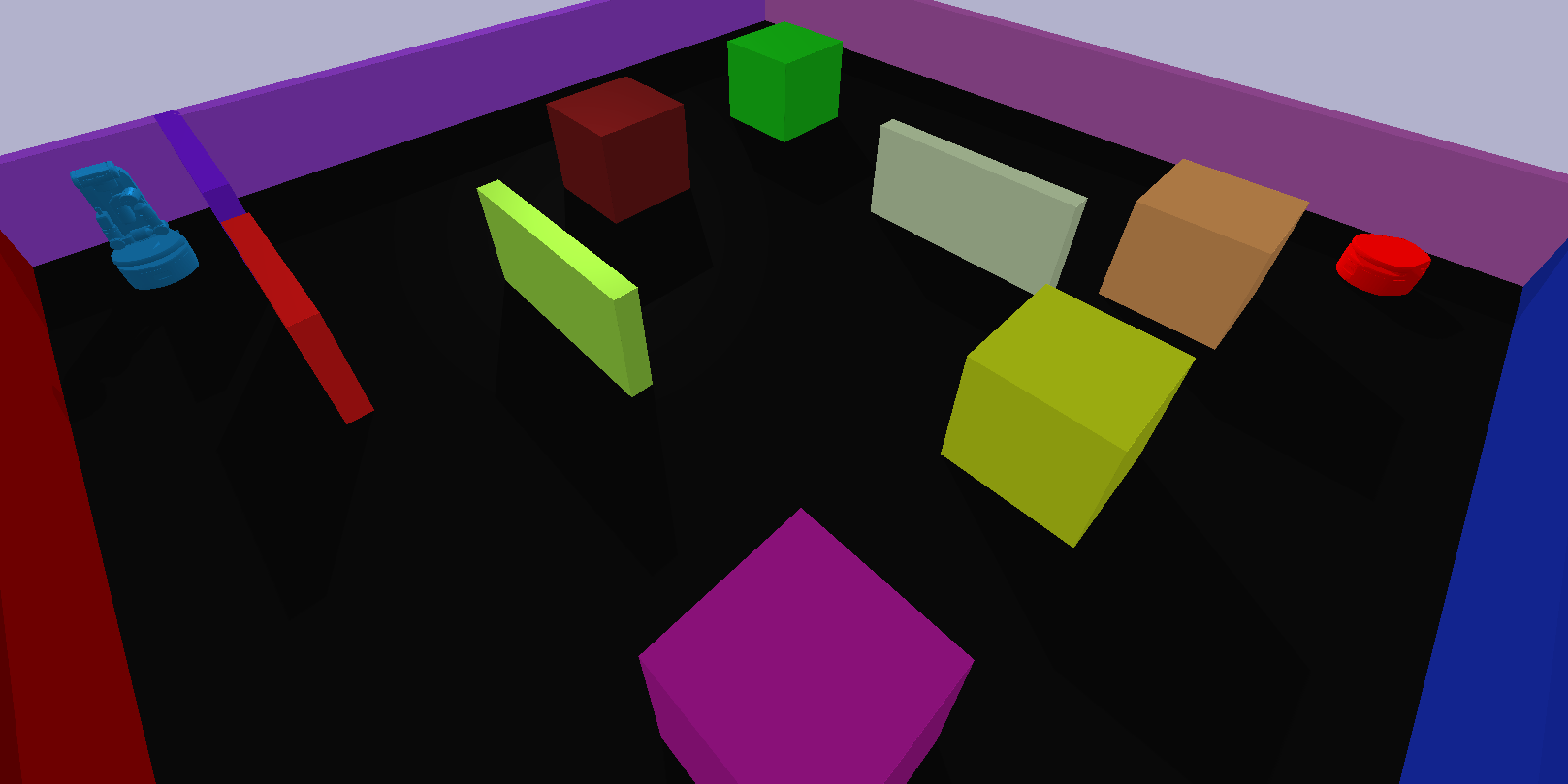}\label{fig:nav_world}}
	\end{tabular}
	\caption{\small Simulated testing environments for the rendezvous task. (a) is a simple environment with no obstacles. (b) is an environment with a wall in between the agents. This environment tests whether agents can converge to one side of the wall (left or right) and thus break the symmetry in the goal alignment problem, a well known multiagent challenge \cite{li2019symmetry,li2020new,chapman2014symmetry}. (c) is a challenging navigation environment with many obstacles in the way of rendezvous.
	\label{fig:worlds}}
\end{figure*}

\begin{figure}[tb]
    \resizebox{1.0\textwidth}{!}{
    \begin{tabular}{@{}cccc@{}}
    \subfloat[Simple world. ]{\includegraphics[height=2.0cm,keepaspectratio=true]{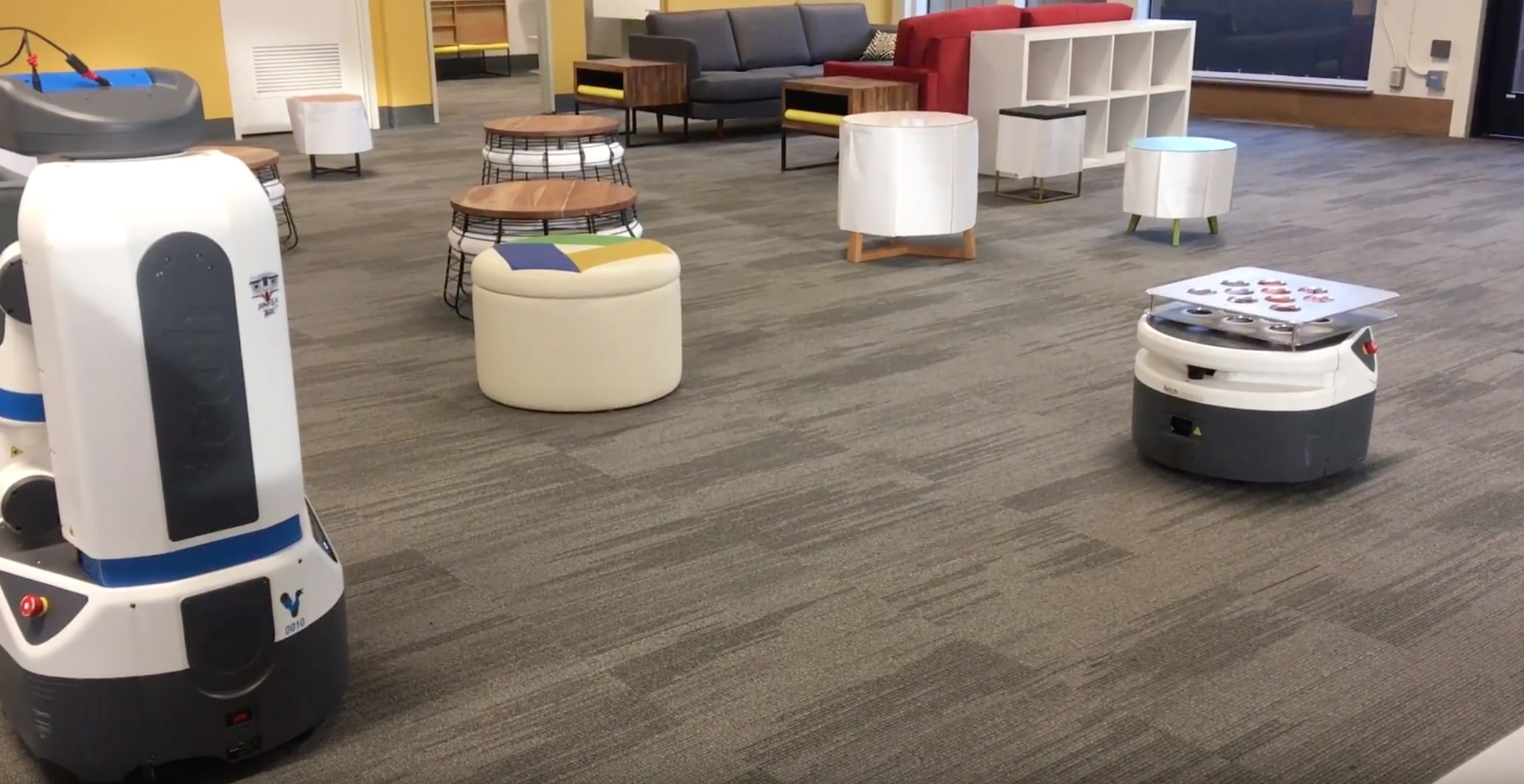}}  & 
    \subfloat[Wall world.]{\includegraphics[height=2.0cm,keepaspectratio=true]{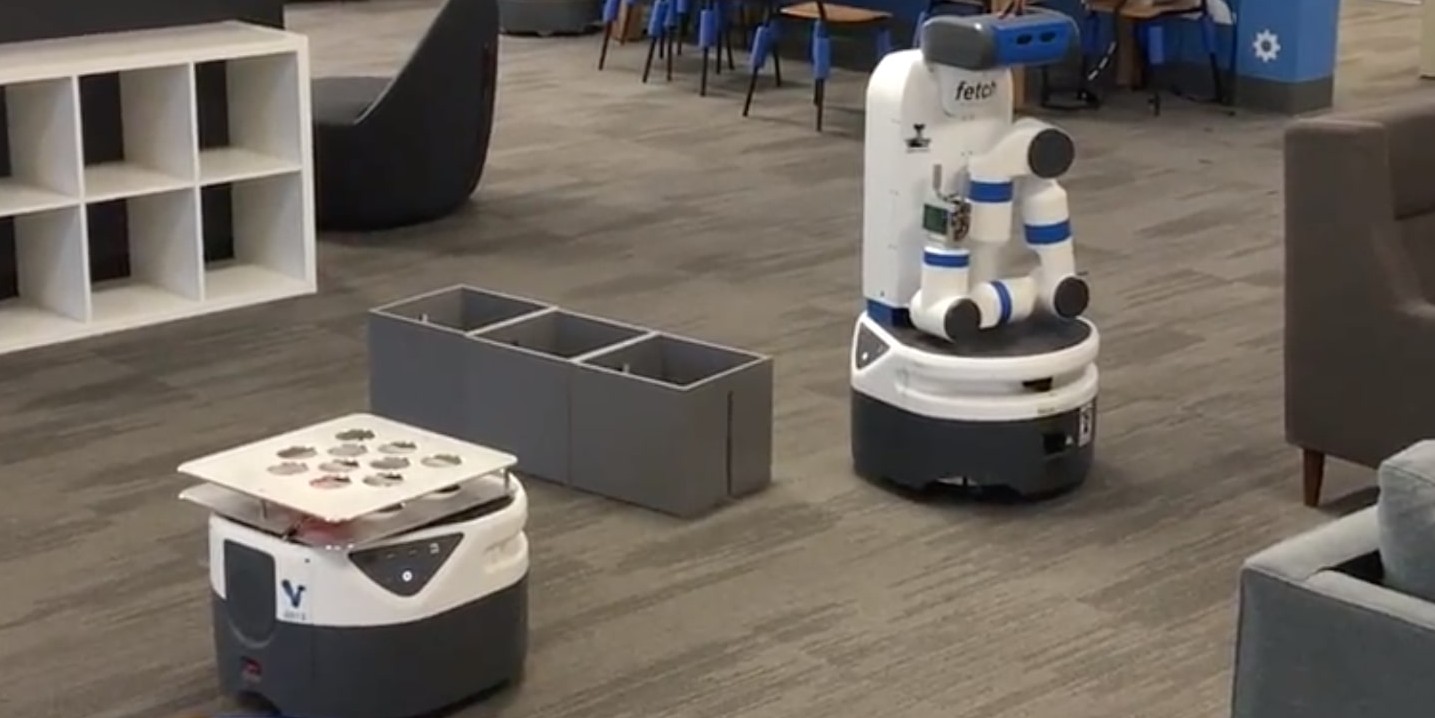}} & 
    \subfloat[Table forest world.]{\includegraphics[height=2.0cm,keepaspectratio=true]{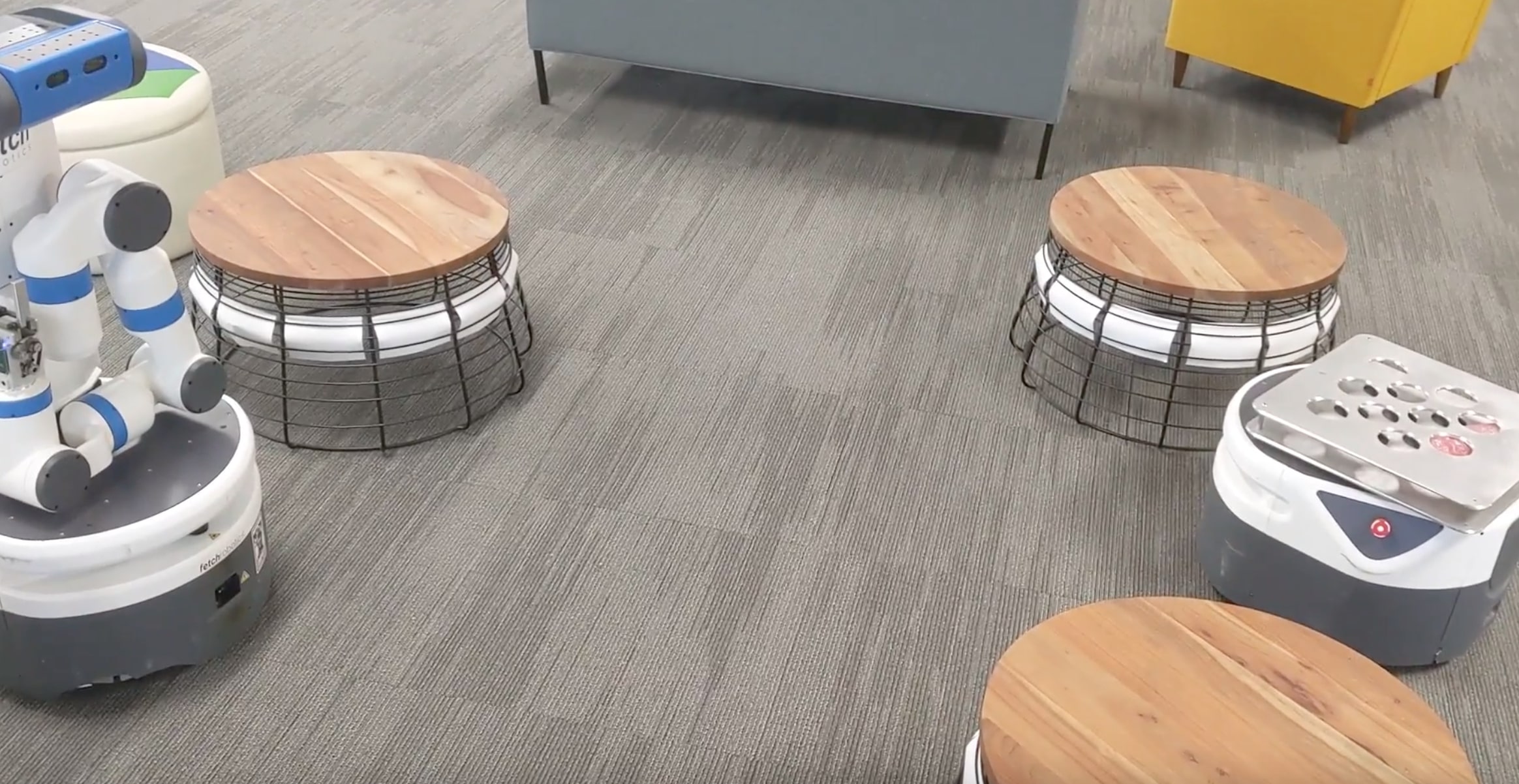}} &
    \subfloat[Roundabout world. ]{\includegraphics[height=2.0cm,keepaspectratio=true]{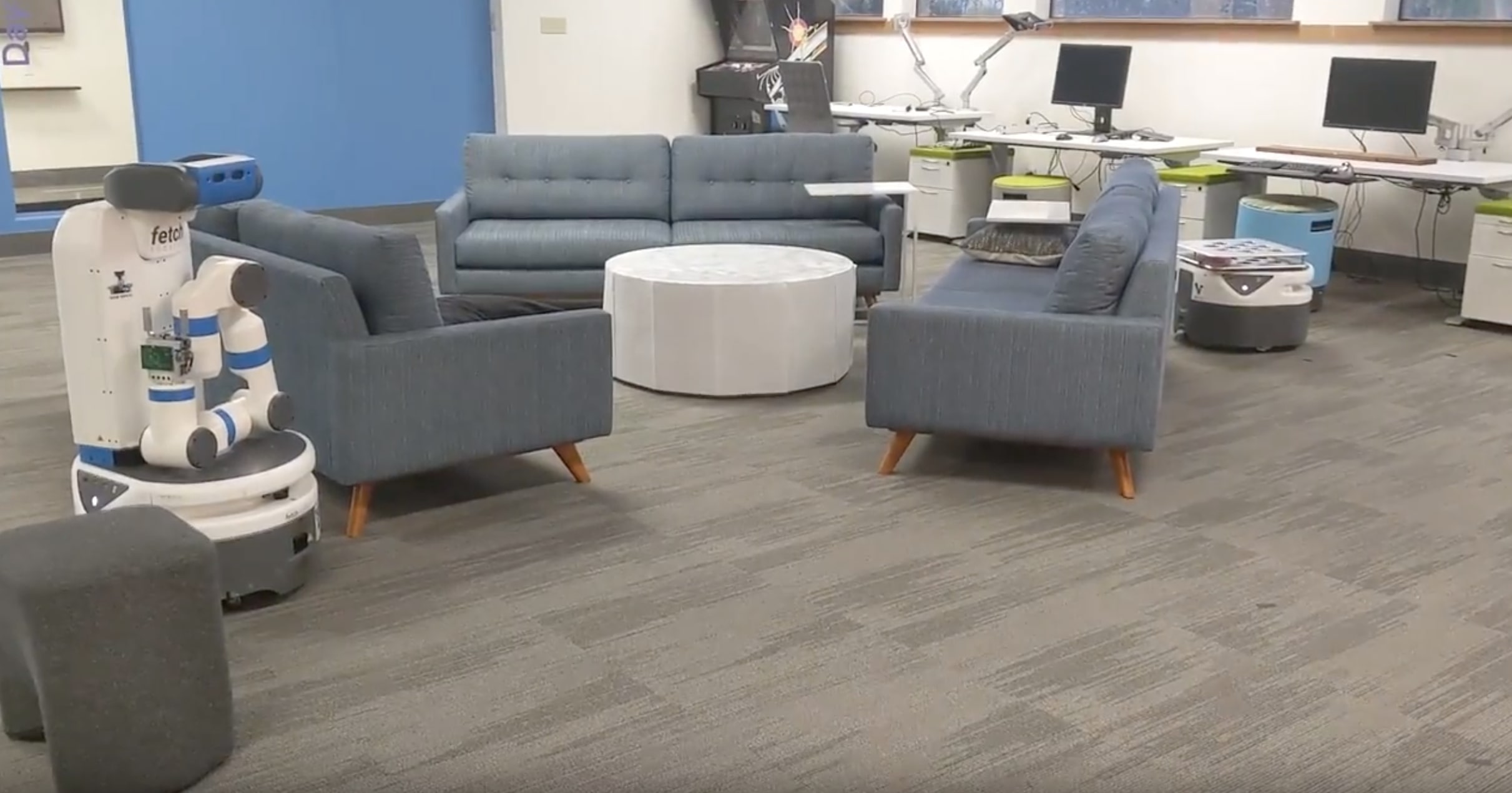}} 
    \end{tabular}}
    \caption{\small Real-world environments used for evaluating \alg\ and \texttt{MADDPG} prior to COVID-19. These environments are also reconstructed from high-fidelity scans in order to compare performances of other baselines.} \label{fig:real_world_envs}
\vspace{-1\baselineskip}
\end{figure}

\textbf{Environments} We use three sets of environments. First, we train the motion prediction models in simulation in a cluttered environment depicted in Figure \ref{fig:training_env}. The training details and learning results are in the Appendix \ref{app:pred_model_setup}. Next, we evaluate \alg\ in two sets of separate environments: three simulation (Figure~\ref{fig:worlds}) and three real environments (Figure~\ref{fig:real_world_envs}). The evaluation environments are designed with different difficulty levels. Due to the COVID-19 shutdowns, we were unable to run all the baselines in the real world. Therefore, we constructed high-fidelity scans (Appendix~\ref{appendix:real_world_videos}) and evaluated all baselines in these environments. We report both results.



\subsection{Leveraging prediction models to align goals \label{sec:testing_results}}

\begingroup
\setlength{\tabcolsep}{3pt}
\begin{figure}[tb]
    \centering
    \begin{tabular}{ccc}
    \subfloat[Simple world evaluation. ]{\includegraphics[width=0.32 \textwidth]{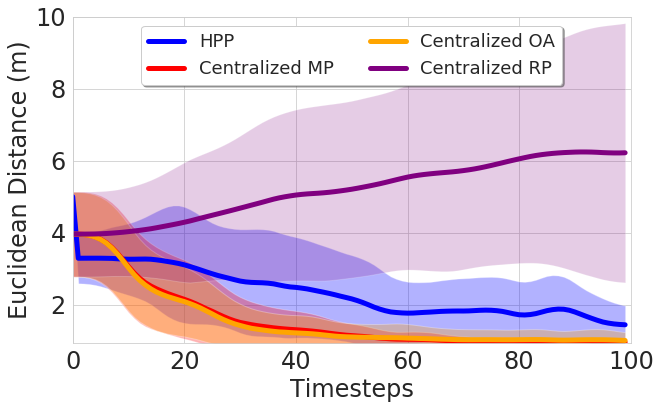}}  & 
    \subfloat[Wall world evaluation.]{\includegraphics[width=0.32 \textwidth]{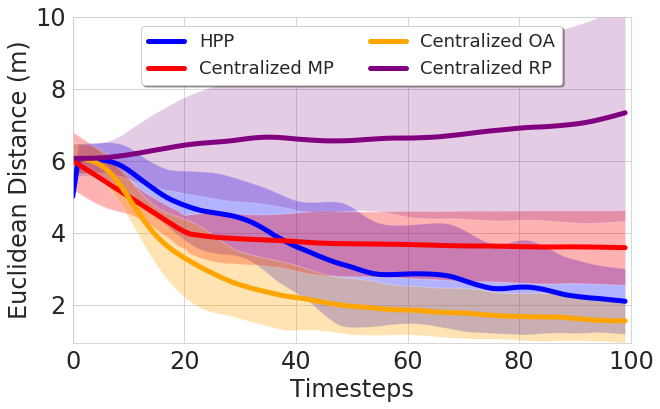}} & 
    \subfloat[Navigation world evaluation.]{\includegraphics[width=0.32 \textwidth]{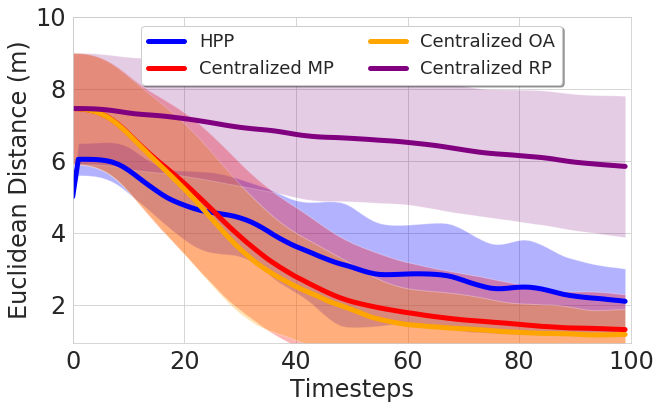}} 
    \end{tabular}
    \caption{\small Distance between agents (lower is better) over planning time in unseen simulation environments (Figure \ref{fig:worlds}). \alg\ (blue) vs centralized baselines. \label{fig:hpp_heuristics}}
\end{figure}
\endgroup


Figure~\ref{fig:hpp_heuristics} demonstrates that \hpp\ effectively leverages prediction models for goal alignment. It compares \hpp\ to centralized heuristic baselines in the simulated test environments (Figure~\ref{fig:worlds}). 
First, \alg\ brings agents closer together than the fixed random goal baseline \texttt{Centralized RP}, illustrating that goals must be both aligned across agents and set intelligently (e.g. set using prediction models or an improved heuristic) for agents to align goals efficiently. Second, \hpp\ leverages its prediction model to actively select better goals than \texttt{Centralized MP} in the wall environment, where \texttt{Centralized MP} is unable to overcome the wall obstacle. 
Furthermore, \hpp\ leverages its prediction model to actively align agent trajectories for the rendezvous, unlike \texttt{Centralized OA}. \href{https://youtu.be/-ydXHUtPzWE}{Visual inspection} of their video performance shows that \texttt{Centralized OA} agents end up \textbf{chasing each other} in tight circles, without stopping, because they're unable to predict where the other agent might be in the future. Although they end up closer together, they do not successfully rendezvous.  


\subsection{Reducing rendezvous time}

\begingroup
\setlength{\tabcolsep}{3pt}
\begin{figure*}[h]
    \centering
    \begin{tabular}{ccc}
    \subfloat[Simple world evaluation. ]{\includegraphics[width=0.32 \textwidth]{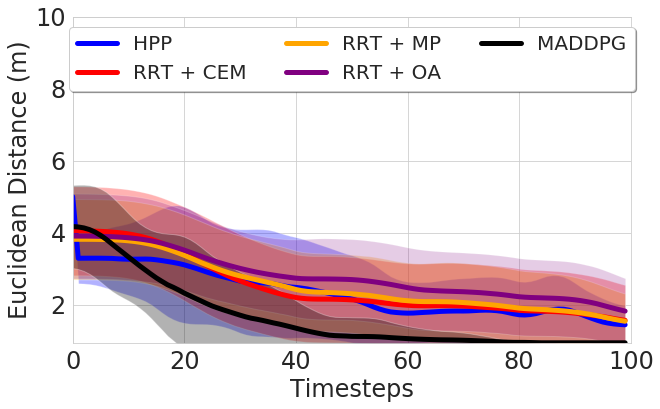}}  & 
    \subfloat[Wall world evaluation.]{\includegraphics[width=0.32 \textwidth]{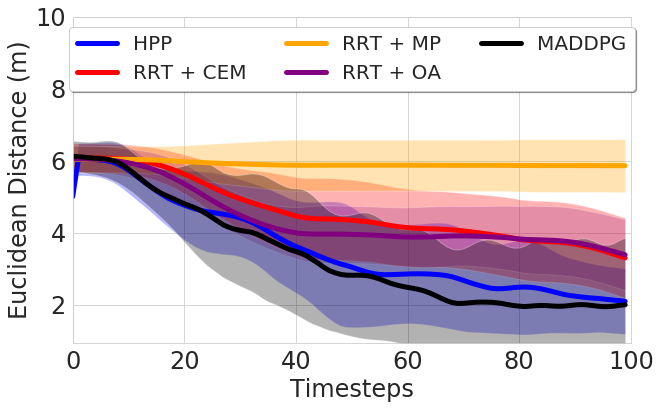}} & 
    \subfloat[Navigation world evaluation.]{\includegraphics[width=0.32 \textwidth]{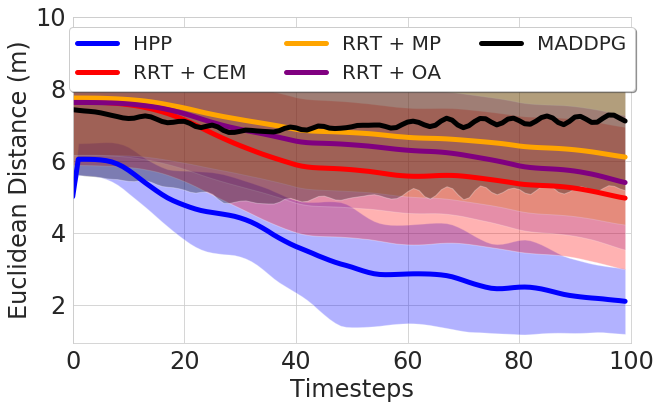}} 
    \end{tabular}
    \caption{\small Distance between agents (lower is better) over planning time in unseen simulation environments (Figure \ref{fig:worlds}). \alg\ (blue) vs. \texttt{RRT}, \texttt{RRT} using heuristics and \texttt{MADDPG}. \label{fig:hpp_rrt_maddpg}}
\end{figure*}
\endgroup

Figure~\ref{fig:hpp_rrt_maddpg} shows that \hpp\ reduces the rendezvous time more than the decentralized baselines---\texttt{RRT+CEM} and \texttt{MADDPG}---and   \texttt{RRT} using centralized heuristics. In \texttt{RRT+OA}, each agent builds and traverses an RRT from its position to the other agent's position. The RRT is recalculated every 40 timesteps. \texttt{RRT+MP} is similar, but the goals of the RRTs are the midpoint between the two agents. 
\alg\ performs similarly to \texttt{RRT} alternatives in the simple environment, and outperforms the baselines in all other environments. Similar to \texttt{Centralized MP}, all three \texttt{RRT} baselines are unable to overcome the symmetry challenge; \texttt{RRT+MP} fails because the \texttt{RRT} planner fails to plan around the wall, \texttt{RRT+OA} fails because the \texttt{RRT} paths are inefficient to overcome the wall obstacle. \texttt{RRT+CEM} performs slightly better than other \texttt{RRT} alternatives in the navigation environment---we hypothesize CEM facilitates aligning the goals of agents. Nonetheless, \texttt{RRT+CEM} performs much worse than \alg\ because the RRT causes frequent backtracking and zigzagging. This inefficient motion makes it difficult for the CEM to distinguish a good goal from a bad based on the agents' early progress toward it.
These failure modes persist in the navigation environment, a more challenging domain.


Finally, we see that \texttt{MADDPG} outperforms \alg\ in the simple world. \texttt{MADDPG} performs similarly \alg\ in the wall world and completely fails in the hard navigation world, where agents only turn in circles. We found it surprising that \texttt{MADDPG} was able to generalize to any testing environments because we expected a model-free multiagent algorithm to require fine-tuning in unseen environments. We hypothesize \texttt{MADDPG} learns to use agent poses and ignores lidar observations. This would explain why it performs poorly in obstacle-laden environments where lidar observations are key in navigation, yet performs similarly to our method in simple, obstacle-minimal environments.

\subsection{Generalization to real environments \label{section:simulated_real_world_results}}

\begingroup
\setlength{\tabcolsep}{1pt}
\begin{figure}[H]
    \centering
    \begin{tabular}{cccc}
    \subfloat[Simple world. ]{\includegraphics[width=0.24 \textwidth]{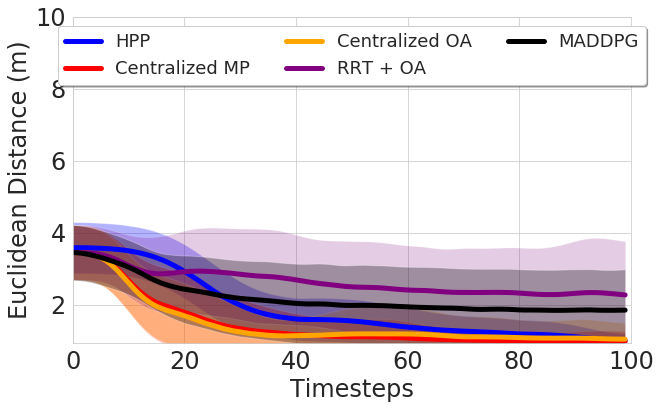}}  & 
    \subfloat[Wall world.]{\includegraphics[width=0.24 \textwidth]{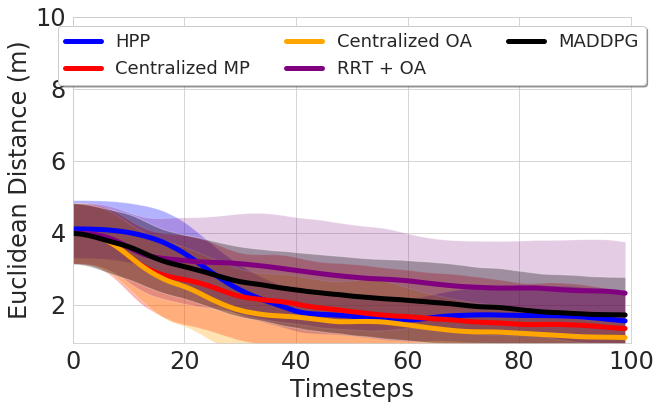}} & 
    \subfloat[Table forest world.]{\includegraphics[width=0.24 \textwidth]{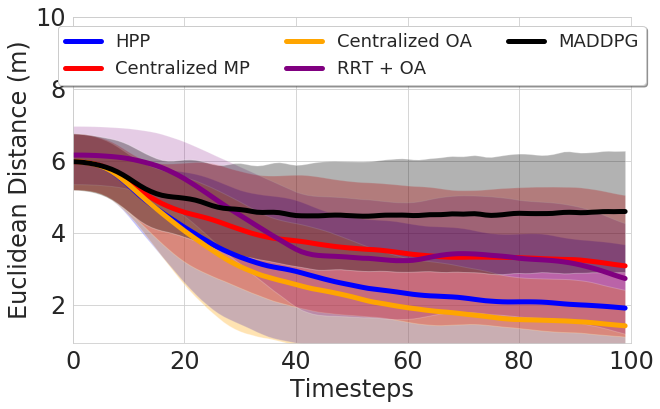}} &
    \subfloat[Roundabout world. ]{\includegraphics[width=0.24 \textwidth]{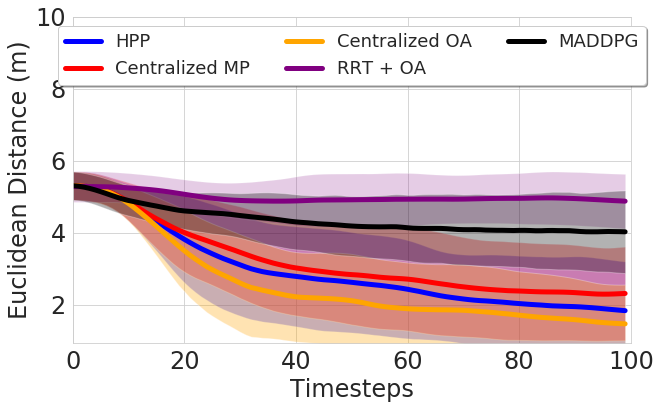}} 
    \end{tabular}
    \caption{\small Distance between agents (lower is better) over planning time in unseen testing environments reconstructed from high-fidelity scans of the real world. \alg\ (blue) vs. \texttt{RRT+OA}, centralized heuristics and \texttt{MADDPG}.
    \label{fig:simulated_real_world_envs_perf}}
\vspace{-1\baselineskip}
\end{figure}
\endgroup

Finally, Figure~\ref{fig:simulated_real_world_envs_perf} shows that \hpp\ generalizes to real environments with zero-shot transfer. 
Appendix~\ref{appendix:real_world_videos} includes the video link to these experiments, images of the reconstructed environments, and some trajectories of the real-world experiments with \alg\ .



The results are consistent with the previous section: in the simple environment, all the baselines perform similarly to each other; in the wall environment, \texttt{Centralized MP} struggles to overcome the symmetry challenge from the wall object; and finally, as environments become more challenging, \texttt{MADDPG} struggles to scale and utilize its lidar observations to navigate, resulting in agents roaming in circles and unsuccessfully converging to a rendezvous location. As before, \texttt{Centralized OA} agents closely chase each other whereas \texttt{HPP} complete the task and handle this type of miscoordination. We hypothesize the reason that \texttt{HPP} struggles on the table top world is due to the narrow navigation spaces in contrast to the open spaces in other environments; the narrowness discourages agents from meeting and aligning goals during their prediction rollouts without colliding into surrounded objects.

\subsection{Environment difficulty and hyperparameters}
Increasing the planning decision making frequency and planning horizon yield disproportionately larger performance improvements in the more complex environments, potentially justifying increased computational cost. We recommend tuning the decision making rate once per environment. See Appendix~\ref{app:pp} for details.

Ablation studies (Appendix~\ref{sec:prediction_model_type}) show that learning to predict change in pose relative to the agent that makes predictions yields better prediction models than learning to predict an absolute pose or change in pose in the global coordinate system. This is reasonable because the predictions are made from the vantage point of the agent making the decisions.  

\subsection{Discussion, future work, and broader impact}
While this work focuses on goal alignment in the context of the navigation \task, \alg\ can be used for other \textbf{decentralized multiagent coordination tasks without explicit coordination} in the future, by applying a different \textit{task objective}, \textit{agent's primitive skills}, and \textit{team size}. First, we assume that all agents share a \textit{task objective}, specified with a reward function in Eq. \eqref{eq:reward}. Substituting a different task reward will result in learning the task defined with the given reward function. Second, this work builds a modular system out of learned components, by assuming pretrained P2P navigation policies are given. Since \alg\ and motion model predictions do not make assumptions about the nature of policies, P2P navigation policies can be substituted with task-appropriate \textit{skill primitives} including \textit{heterogeneous teams of robots}. Lastly, the algorithm makes no assumptions about the \textit{team size}, and future work can consider larger teams.


%

\section{Conclusions}

This work presents a model-based RL approach to solving a decentralized \task.
First, the method learns to approximate own and teammates' motion models that reflect robots' abilities and limitations with respect to their control policies via self-supervised learning; the motion models decouple the capabilities of the agents from the task. 
Next, the method uses a planning module that iteratively proposes subgoals for the agent to move towards. 
\hpp\ maintains a belief distribution over goals using a joint objective function and the learned motion models. Finally, \hpp\ updates this distribution and replans using motion predictions to address misaligned agent goals, i.e. miscoordination.
We demonstrate the generalization of the motion predictor models to new environments, both in simulation and in the real world. Compared to the baselines, without explicit coordination between the agents, HPP is more likely to complete the \task\ by selecting goals that are feasible, coordinated, and unobstructed.
\acknowledgments{

We thank Michael Everett, Oscar Ramirez and Igor Mordatch for the insightful discussions.
}

\bibliography{example}  

\newpage

\appendix
\section*{Appendix}
\section{Discussion on the \textit{rendezvous} task reward \label{app:reward}}
Our work uses a reward function that captures the negative distance between agents’ final simulated positions,

\begin{equation}
     \reward(\pose_{1..n}) = 
    \begin{cases}
        0 & |\pose_{j} - \pose_\mu| < \goaldistance \quad \forall j \in 1..n \\
        \sum_{j, k\neq j } -|\pose_k  - \pose_j| & otherwise
    \end{cases}
    \label{eq:reward1}
\end{equation}
where $\pose_\mu = \frac{1}{\numagents}\sum_{k \in 1..n} \pose_k$ is the center point of the robots' positions (average of their locations), $\goaldistance$ controls the precision of the rendezvous, $\numagents$ is the number of agents.

A previous reward function we used was the negative accumulated distance over a simulated trajectory. However, this assumes agents to strictly close the distance between themselves at every step. In experimentation, we realized agents cannot strictly minimize the distance at every step and selected suboptimal goals with the previous reward function. For example, in obstacle-filled environments, agents need to first navigate around obstacles before rendezvousing. The current reward is designed based on this insight: agents should only care about their final positions, not intermediate ones.

\section{Evaluation setup \label{appendix:setup}}

\subsection{Robot and planning parameter setup}
\label{app:robot_setup}
We use two robots, a Fetch and a Freight \cite{wise2016fetch}. Each observes a 222 beam 2D lidar with 220 degree field of view to match the real robot sensors, as well as the global poses of all agents.
Each agent's action space is linear and angular base velocity, clipped to the ranges $[0, 1]$ meters/second and $[-3, 3]$ radians/second respectively. Linear acceleration is limited to $0.4\ m/s^2$ and angular acceleration to $1.48\ \text{rad}/s^2$ ($85^{\circ}/s^2$). In our experiments, the agent's low-level policy operates every timestep (i.e. $\lowRate = 1\,$) and the high-level policy operates every 10 timesteps (i.e. $\highRate = 10\,$), unless otherwise specified. The rollout length $T$ is 5. The episodes are 100 $\lowRate$ timesteps (20 s) long both in training and in evaluation. In the real world experiments, the ROS navigation stack \cite{288} provides the pose observations. As for CEM, $\epsilon = 0.001$, the number of max sampling iterations is 15, the number of samples is 15 and the number of top samples used to update the goal distribution is 5. For the low-level controller, we use a goal-conditioned navigation policy $\ptwop_i$ trained with \cite{autorl}, although any other P2P policy capable of driving a robot to a goal using local observations would be suitable, such as \cite{unstuck-dinesh}.

\subsection{\texttt{MADDPG} setup}
\label{app:maddpg}

\begin{figure}[tbh]
    \centering
    \includegraphics[width=0.35 \textwidth]{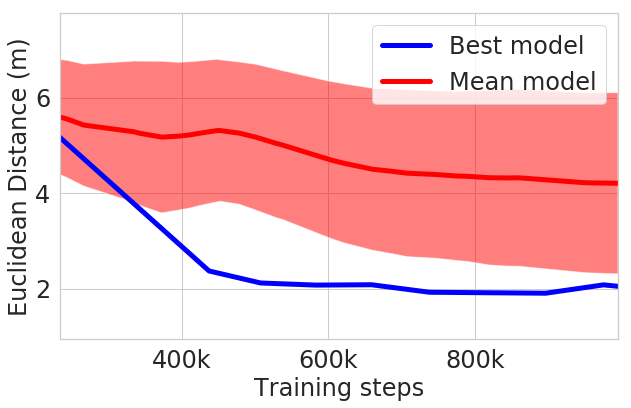}
    \caption{\small MADDPG training curves. The pink shaded region shows the standard deviation across a batch of training runs. The blue curve is the best trained MADDPG model, which we use as the baseline. \label{fig:maddpg-train}}
\end{figure}

Both the policy and critic networks for \texttt{MADDPG} are two layer networks with 64 units in each. Batch size is 1024 and learning rate is 0.0005. \texttt{MADDPG} was trained on 10 different seeds in the randomized obstacle environment described in Section~\ref{sec:methods}. \texttt{MADDPG} takes 1M steps to converge (Fig.~\ref{fig:maddpg-train}). \texttt{MADDPG} is also relatively unstable; of 10 policies, 3 failed to converge, while the prediction models are stable to train. In the experiments for Section~\ref{section:experiments}, we used the policy that performed the best (i.e. achieved the highest reward) in the \task.

\begin{figure}[tb]
    \centering
    \includegraphics[width=0.4 \textwidth]{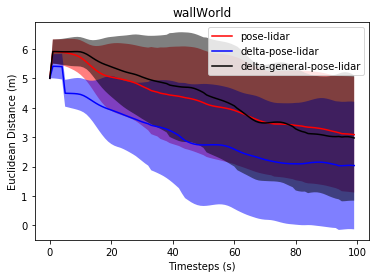}
    \caption{\small Performance of \alg\ on wall world varying the prediction model. Lower is better. \texttt{delta-pose-lidar} (blue) is ours. \label{fig:prediction_perf}}
\end{figure}

\subsection{Setup for prediction models}
\label{app:pred_model_setup}
Figure~\ref{fig:training_env} shows the training environment for the predictive models $\f_i$ and $\f_{-i}$ and Figure~\ref{fig:worlds} three previously \textit{unseen} test environments. The simulated evaluations are repeated 10 times where the agents are randomly initialized 5 meters from each other. 

The prediction model is a 4-layer fully connected network with layer units 64, 128, 128 and 64 and a learning rate of 0.001. It is trained on experiences collected from an environment with randomized furniture obstacles (Figure~\ref{fig:training_env}). Goals are randomized within a $20 \times 20$ meter square and are not guaranteed to be collision-free. The history trace length, $H$, is 5 for all agents. 
Low-level policies navigating towards randomly selected goals collect 50,000 trajectories, or the equivalent of about 11 days of real-time experience. Each prediction model ($f_i, f_{-i}$) has its own network, trained on 50,000 epochs with batch sizes of 500. The data collection and training take about 3 hours to complete. Both \textit{self-prediction} and \textit{other-prediction} models converge (Figure~\ref{fig:loss}). 
As expected, \textit{self-prediction} is easier than predicting another agent's motion without having its sensor readings.

\section{Ablation studies \label{app:planning_params}}

\subsection{Planning hyperparameters}
\label{app:pp}
We investigate the role of three planning hyperparameters in Algorithm~\ref{high_level_policy}: planning frequency ($\highRate$), planning horizon ($T$), and goal sampling parameters.

\begin{figure*}[tb]
    \centering
    \begin{tabular}{ccc}
    \subfloat[Simple world evaluation.]{\includegraphics[width=0.3 \textwidth]{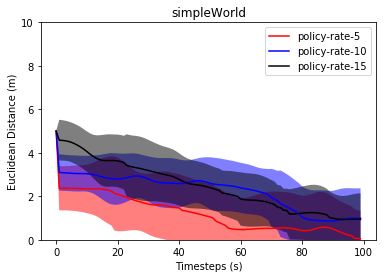}} &  
    \subfloat[Wall world evaluation.]{\includegraphics[width=0.3 \textwidth]{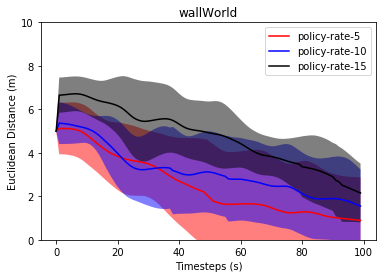}} & 
    \subfloat[Navigation world evaluation.]{\includegraphics[width=0.3 \textwidth]{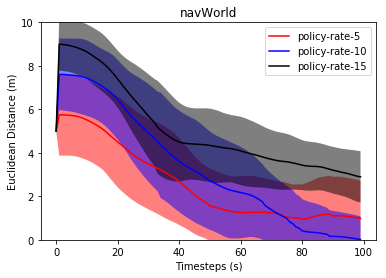}}
    \end{tabular}
    \caption{\small Planning frequency at which the high level policy $\Pi$ recalculates new goal across environments. Lower is better. The planning frequency should be tuned per environment.\label{fig:policy_rate}}
\end{figure*}

\begin{figure*}[tb]
    \centering
    \begin{tabular}{ccc}
\subfloat[Simple world evaluation.]{\includegraphics[width=0.3 \textwidth]{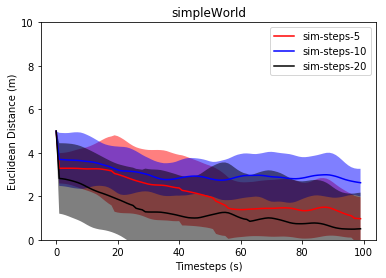}} &  
    \subfloat[Wall world evaluation.]{\includegraphics[width=0.3 \textwidth]{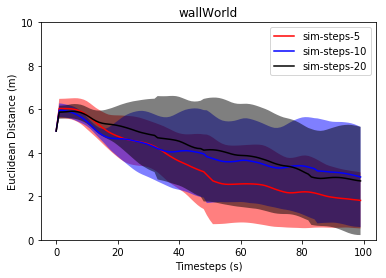}} & 
    \subfloat[Navigation world evaluation.]{\includegraphics[width=0.3 \textwidth]{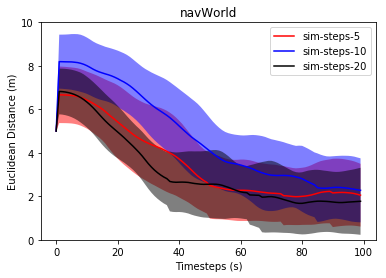}}
    \end{tabular}
    \caption{\small Planning horizon evaluations across environments. Lower is better. Planning horizon does not depend on the environment, and should be tuned once per predictor.\label{fig:nav_traj_length}}
\end{figure*}



\paragraph{Planning frequency} Across the three environments, planning every 5 timesteps leads to faster task completion compared to every 10 or 15 timesteps (Figure~\ref{fig:policy_rate}). Though this is unsurprising, we also note that planning frequency makes more difference in more complex environments. However, frequent planning comes at the  cost of higher computational load, so it is helpful to tune it to the desired precision. Figure~\ref{fig:policy_rate} suggests that the optimal planning rate depends on the environment, so planning rate should be tuned once per environment.

\paragraph{Planning horizon}  Similarly, evidence in Fig. \ref{fig:nav_traj_length} suggests that the performance gap varies based on the environment and the planning horizon. However, the tradeoff of having a longer or shorter horizon is not as significant as varying the planning frequency since an additional rollout is much faster than frequently re-planning and re-evaluating goals, which is what the planning frequency controls.

\paragraph{Goal sampling parameters}
We iteratively improve the sampled goal using CEM, but any alternative importance sampling technique may work as well. We conducted experiments where we varied the parameters of the CEM: maximum number of iterations and number of top goals (between 5 and 15) used for updating the CEM distribution. None of these ablation experiments indicated that \alg\ has a strong dependency on the goal sampling parameters.

\subsection{Prediction models ablation \label{sec:prediction_model_type}}
\begin{figure*}[tb]
    \centering
    \begin{tabular}{ccc}
    \subfloat[Loss for \texttt{pose-lidar}.]{\includegraphics[width=0.32 \textwidth]{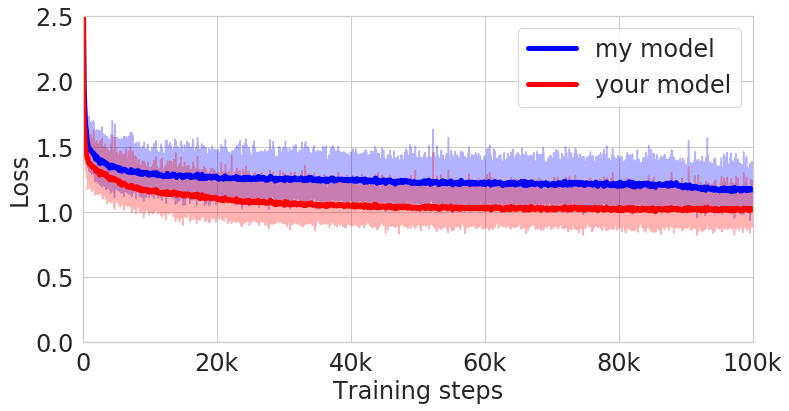}} &  
    \subfloat[Loss for \texttt{delta-pose-lidar}.]{\includegraphics[width=0.32 \textwidth]{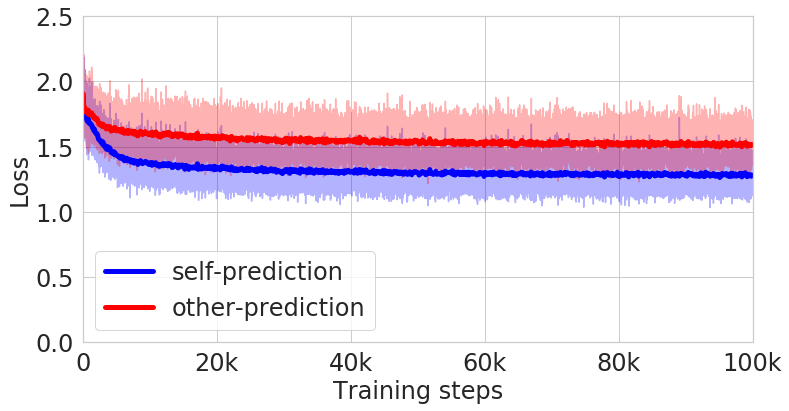}} & 
    \subfloat[Loss for \texttt{delta-general-pose-lidar}.]{\includegraphics[width=0.32 \textwidth]{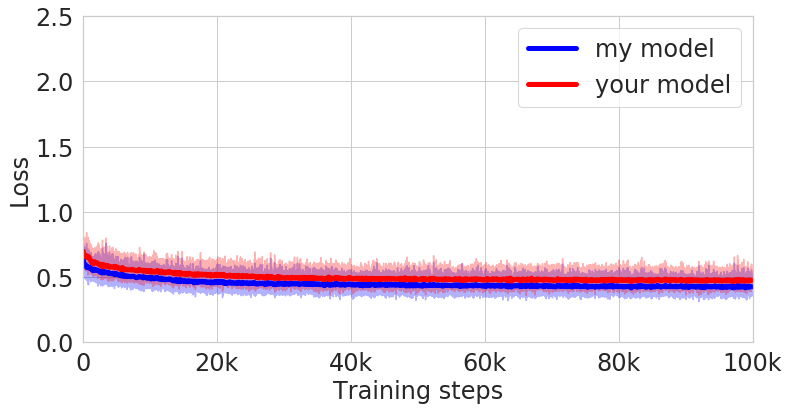}}
    \end{tabular}
    \caption{\small Training losses for the different prediction model types. The dark lines are averaged over 100 steps. \label{fig:loss}}
\end{figure*}

We conduct an ablation study on model representations for $\f_i, \f_{-i}$. 
To contrast prediction methods, we train two additional prediction types under the same settings and inputs. We refer to the prediction type presented in  Section~\ref{section:prediction-models} as \texttt{delta-pose-lidar}. The two additional ones are \texttt{pose-lidar} and \texttt{delta-general-pose-lidar}.
\texttt{pose-lidar} directly outputs the next pose and lidar predictions. 
\texttt{delta-general-pose-lidar} differs from \texttt{delta-pose-lidar} in that it predicts the pose of agent $-i$ relative to the current pose of agent $i,$ and predicts the full lidar observation of agent $i.$ Figure~\ref{fig:loss} shows the training loss for the three prediction types.

We evaluate the task performance of the prediction models on wall world. Fig.~\ref{fig:prediction_perf} shows that \texttt{delta-pose-lidar} performs the best while the other two prediction models perform equally well. In particular, the first few timesteps indicate a drastic decrease in euclidean distance since these agents quickly overcome the symmetry-breaking problem presented in wall world. Future work can investigate different prediction schemes. 


\section{Real-world videos \label{appendix:real_world_videos}}
Our \href{https://youtu.be/-ydXHUtPzWE}{work's video linked here}\footnote{Video link: https://youtu.be/-ydXHUtPzWE} shows that evaluation by the euclidean metric does not tell the whole story: The video illustrates how our model-based reinforcement learning is key for agents to rendezvous. With \texttt{Centralized OA}, agents end up \textbf{chasing each other} in tight circles, which reduces the distance between the agents but is not a successful rendezvous. Even though euclidean distance is a standard way to evaluate success rates in robotics and motion planning, the video shows a discrepancy in how the task is completed by the different methods. 

Thus, the main takeaways from the video are: \texttt{HPP} outperforms the other baselines and can successfully transfer zero-shot sim-to-real; \texttt{Centralized OA} agents do not rendezvous because they chase each other; \texttt{RRT+OA} is unable to complete the rendezvous in the specified episode time because agents plan zigzagging paths which undo their rendezvous progress; and \texttt{MADDPG} agents rarely rendezvous and more often spin in place because the baseline is unable to transfer into real-world environments and handle the noisy, high-dimensional sensor data. 


Some methods are compared in the real world (pre-COVID), and all methods are compared in the simulation environments reconstructed from scans of the same room used in the real world. Images of \alg\ performing in the real world can be seen in Table~\ref{tab:real-world-images}. The robots are decentralized, which means they plan and execute independently. There is no coordination in their action execution and there is no centralized controller telling the robots where to go. 

During the pauses in \alg\ video evaluations, \alg\ samples and evaluates goals in real-time before moving towards the goal of its choosing. 

The videos also include clips of the agents' goals dynamically changing and an example episode illustrating the training environments for the prediction models. In the video, the waypoints sometimes are placed on top of obstacles. This is because agents select waypoints by whichever brings agents closer. Our algorithm doesn’t assume the goals to be reached at the end of a simulated trajectory (Algorithm \ref{high_level_policy}). 




\begin{figure*}[tb]
    \centering
    \begin{tabular}{cccc}
    \subfloat[Simple world. ]{\includegraphics[width=0.24 \textwidth]{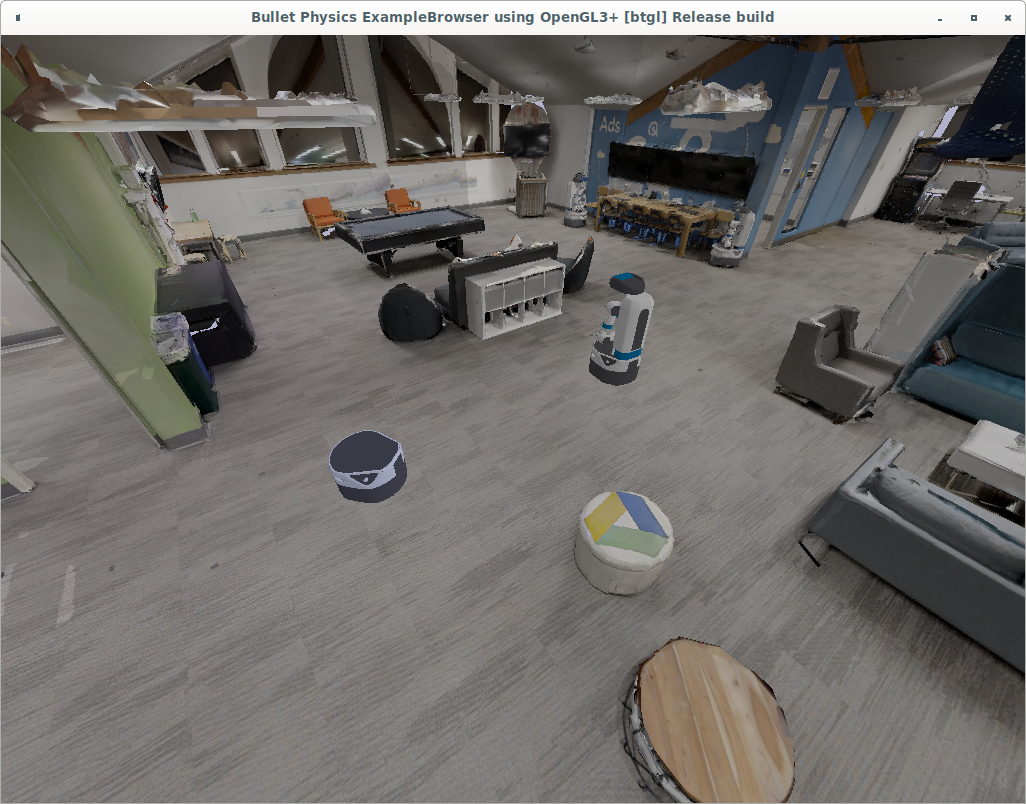}}  & 
    \subfloat[Wall world.]{\includegraphics[width=0.24 \textwidth]{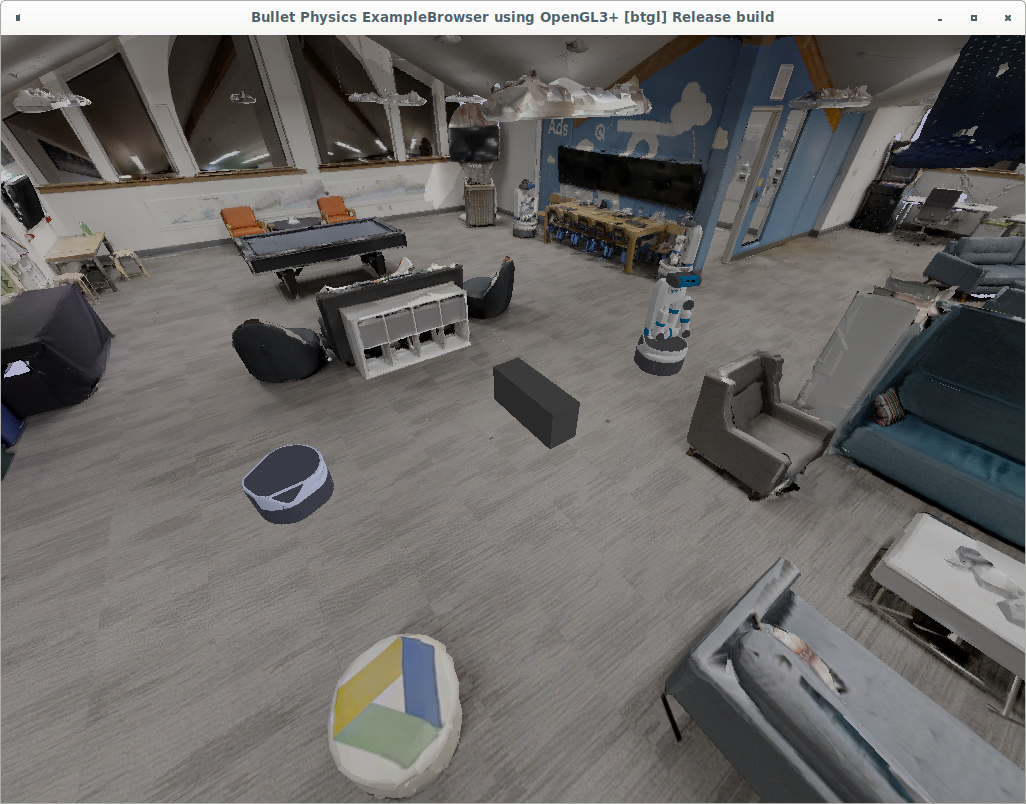}} & 
    \subfloat[Table forest world.]{\includegraphics[width=0.24 \textwidth]{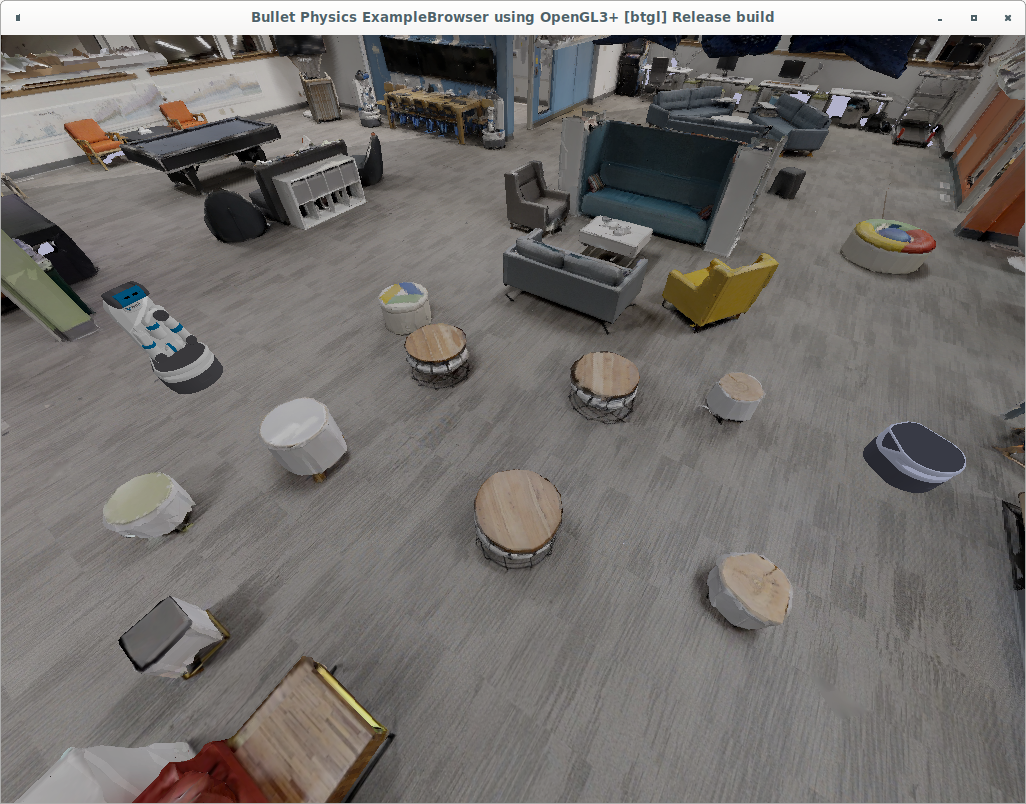}} &
    \subfloat[Roundabout world. ]{\includegraphics[width=0.24 \textwidth]{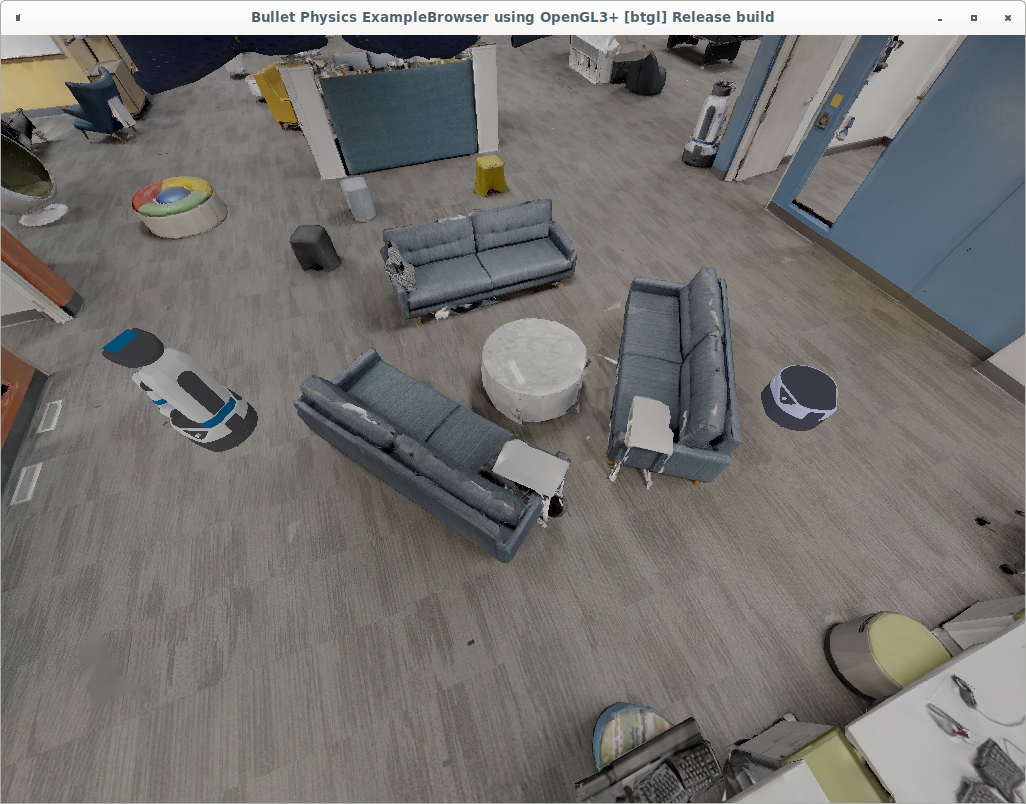}} 
    \end{tabular}
    \caption{ Environments built from scans of the room used for real world evaluations (i.e. Figure \ref{fig:real_world_envs}). These environments were used to evaluate \alg\ and all baselines. \label{fig:simulated_real_world_envs_pics}}
\end{figure*}

\begin{table*}
  [ht]
  \begin{tabular}
      {lllll} \hline Environment type & 1 & 2 & 3 & 4 \\
      \hline \textbf{Simple} & \parbox[c]{0.18\linewidth}{
      \includegraphics[width=1in,height=.6in]{images/real-robots2/simple1.jpg}} & \parbox[c]{0.18\linewidth}{
      \includegraphics[width=1in,height=.6in]{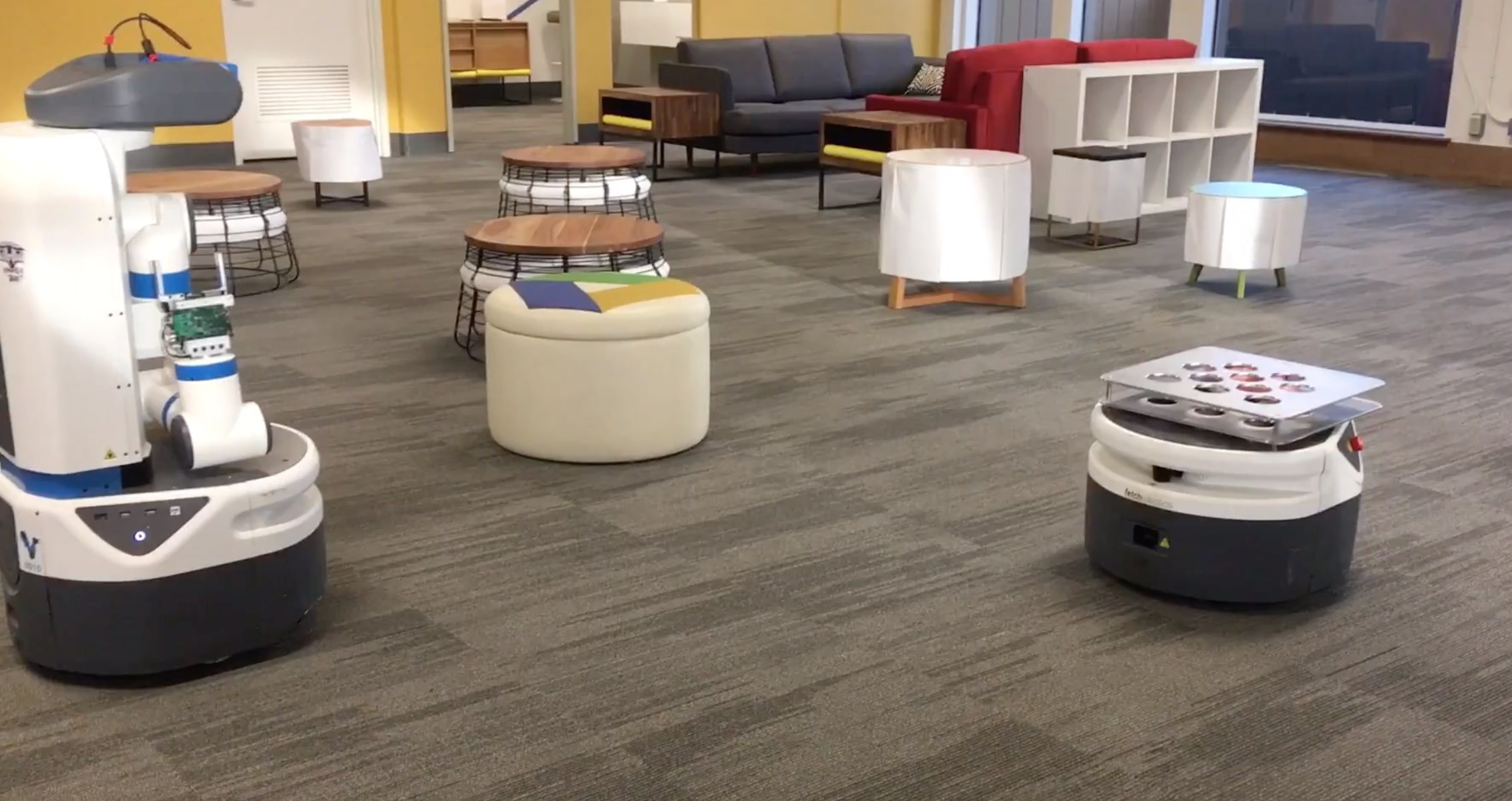}} & \parbox[c]{0.18\linewidth}{
      \includegraphics[width=1in,height=.6in]{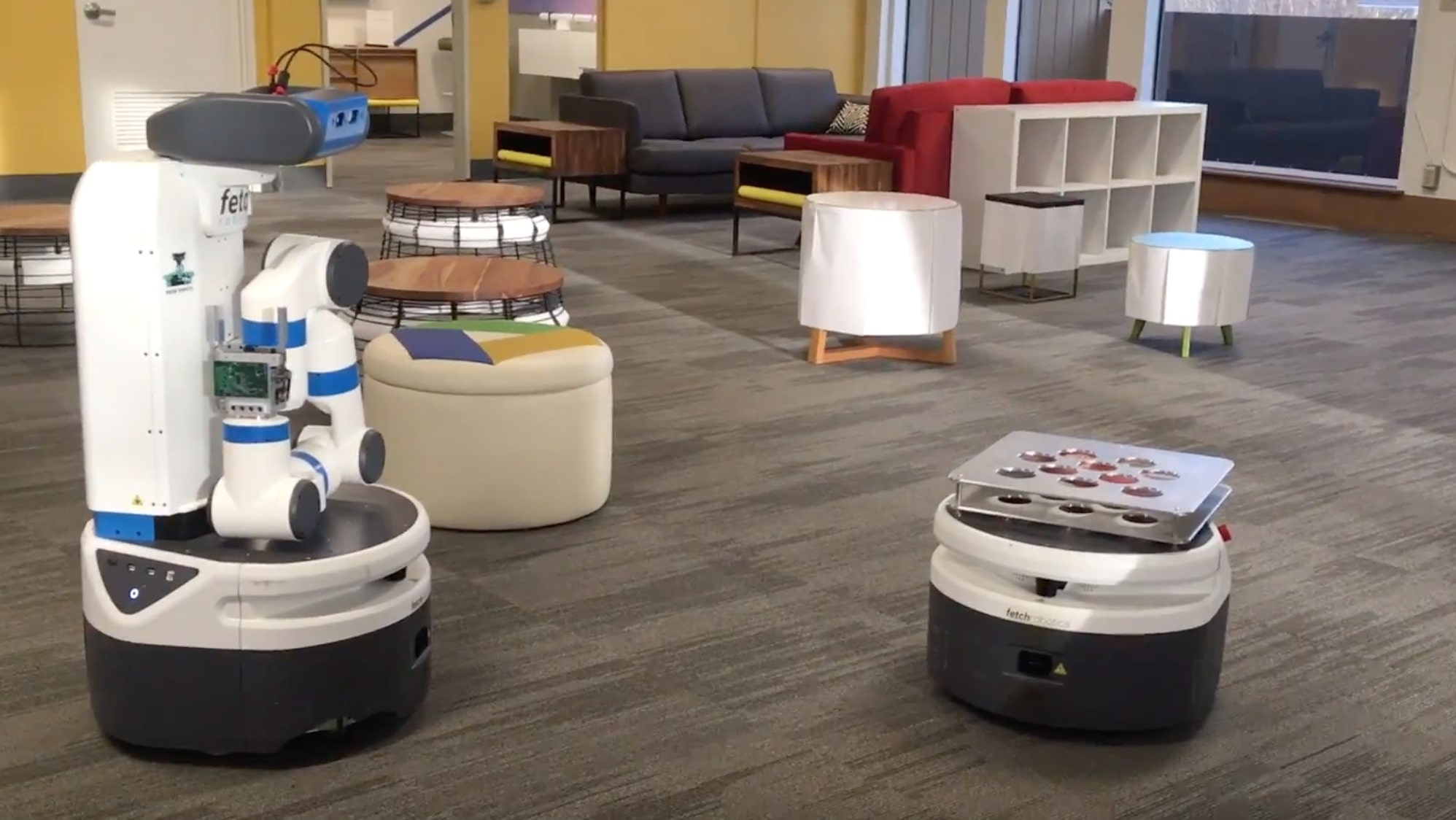}} & \parbox[c]{0.18\linewidth}{
      \includegraphics[width=1in,height=.6in]{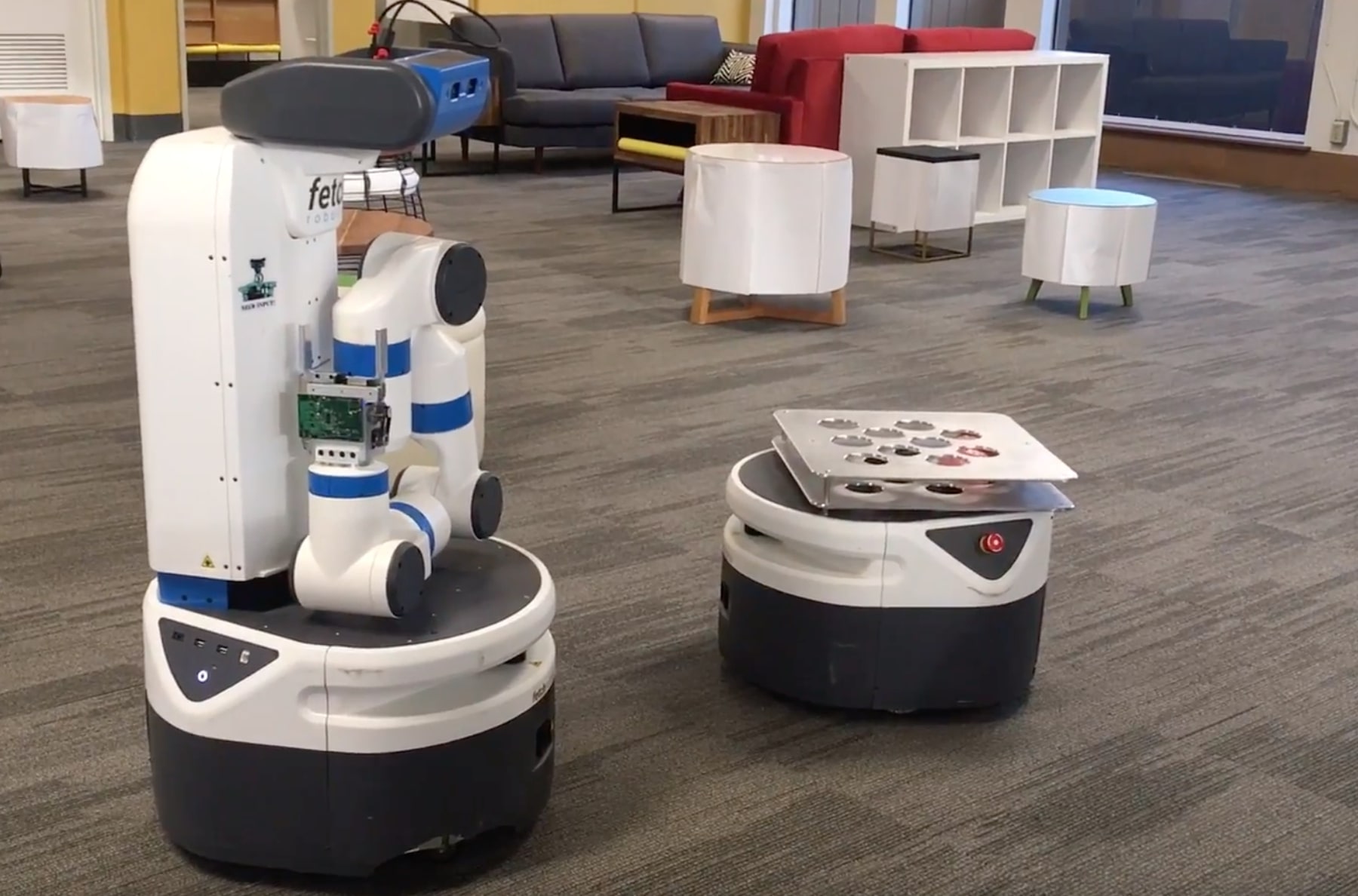}} \\
      \hline \textbf{Wall} & \parbox[c]{0.18\linewidth}{
      \includegraphics[width=1in,height=.6in]{images/real-robots2/wall1.jpg}} & \parbox[c]{0.18\linewidth}{
      \includegraphics[width=1in,height=.6in]{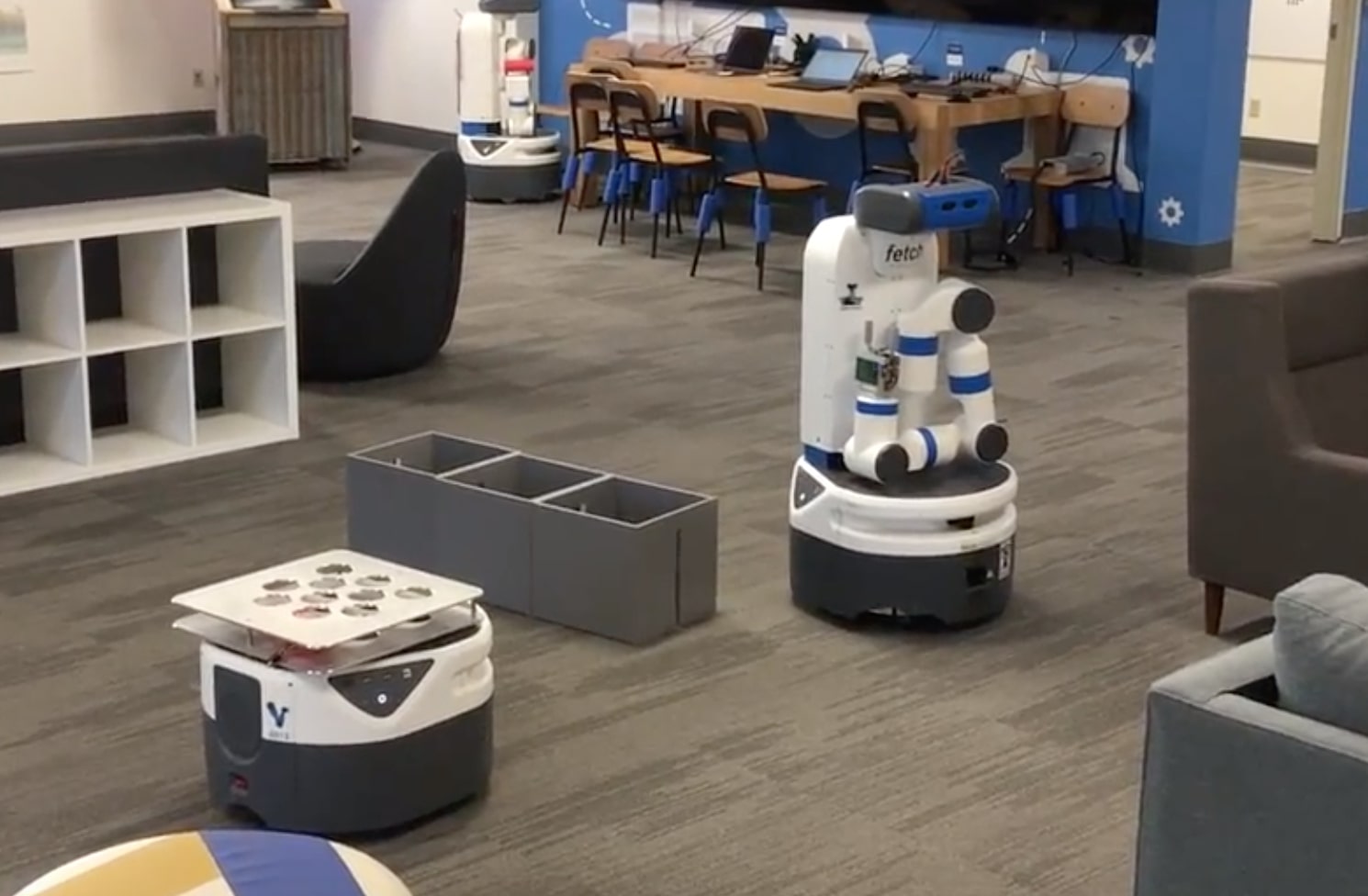}} & \parbox[c]{0.18\linewidth}{
      \includegraphics[width=1in,height=.6in]{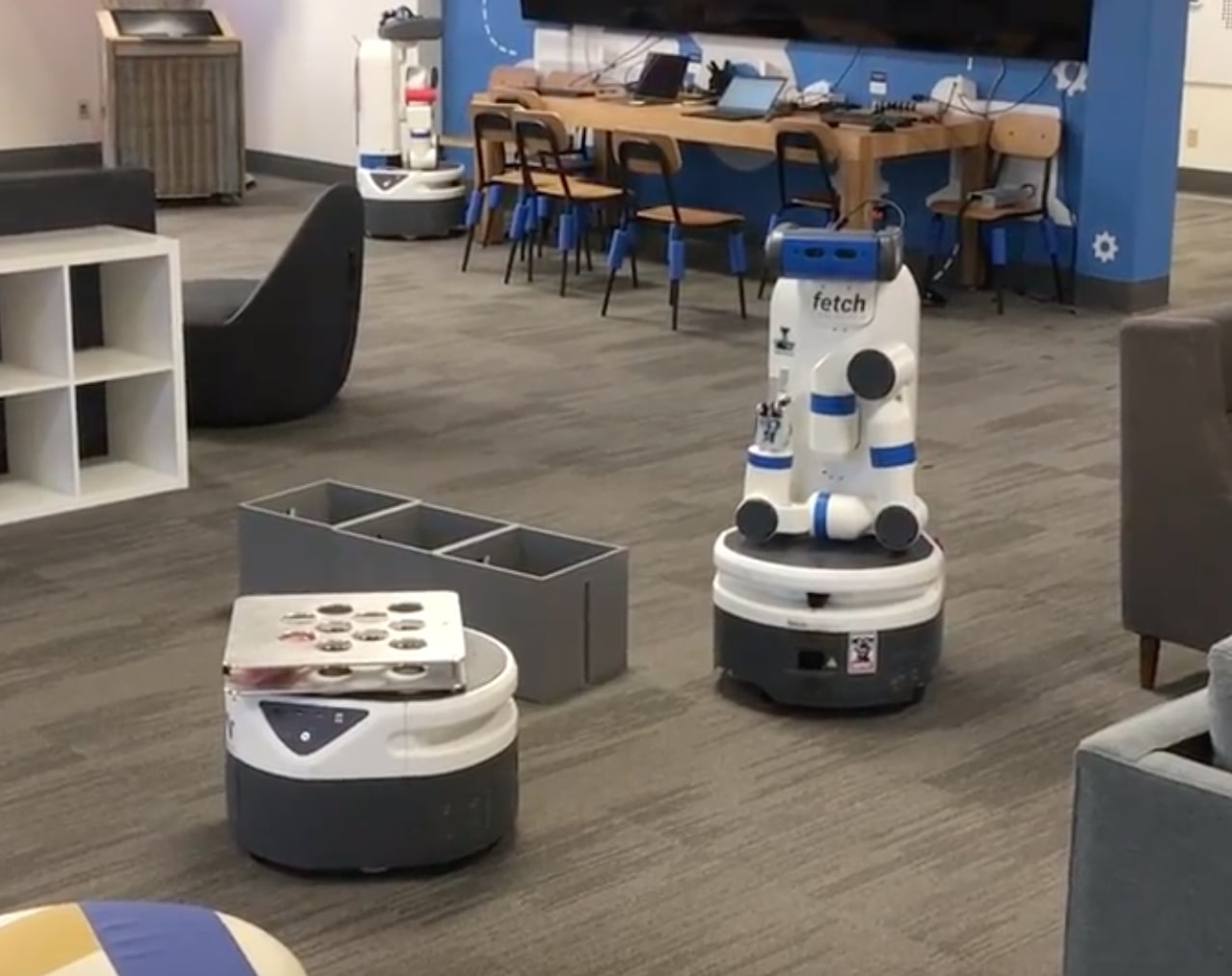}} & \parbox[c]{0.18\linewidth}{
      \includegraphics[width=1in,height=.6in]{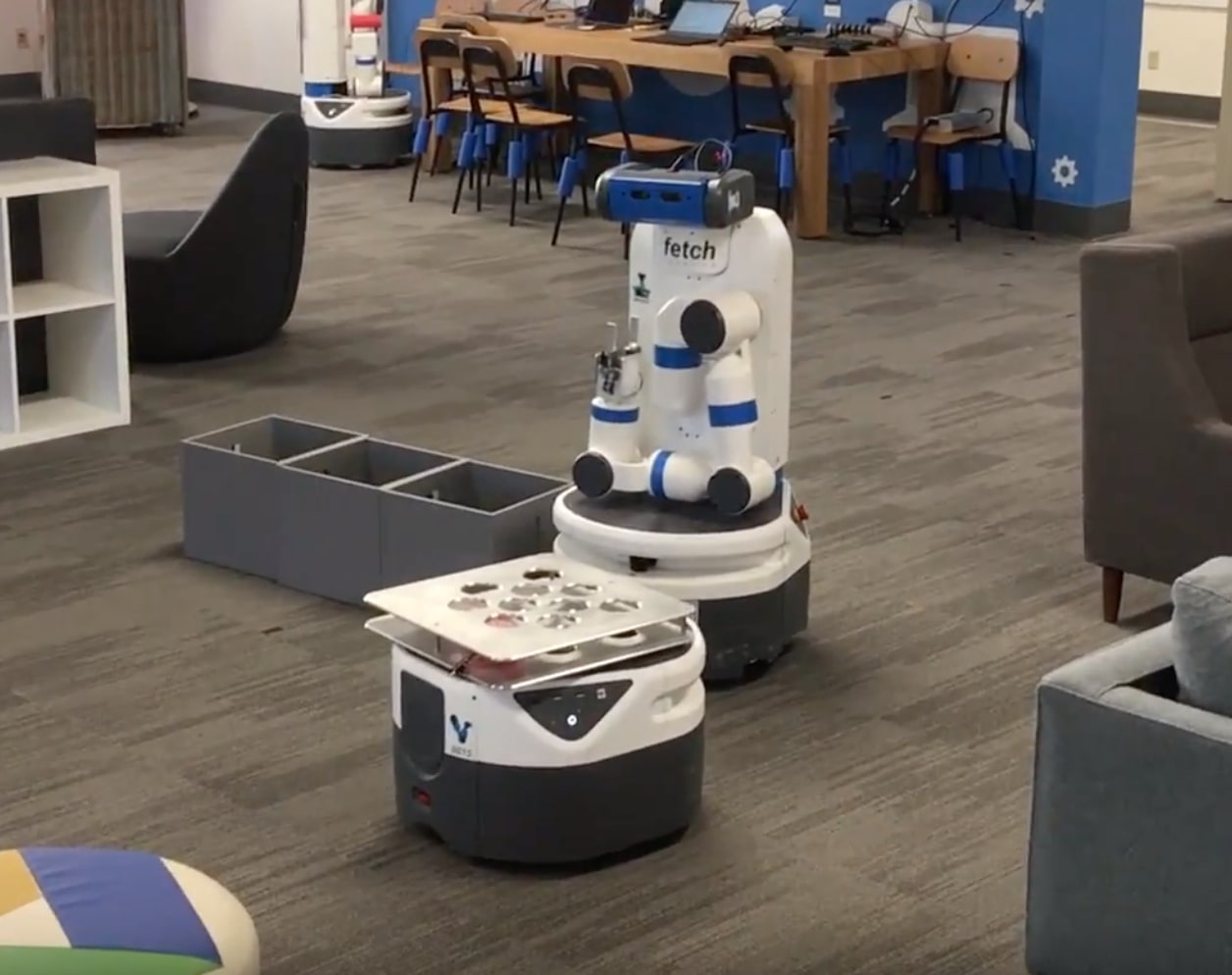}} \\
      \hline \textbf{Forest} & \parbox[c]{0.18\linewidth}{
      \includegraphics[width=1in,height=.6in]{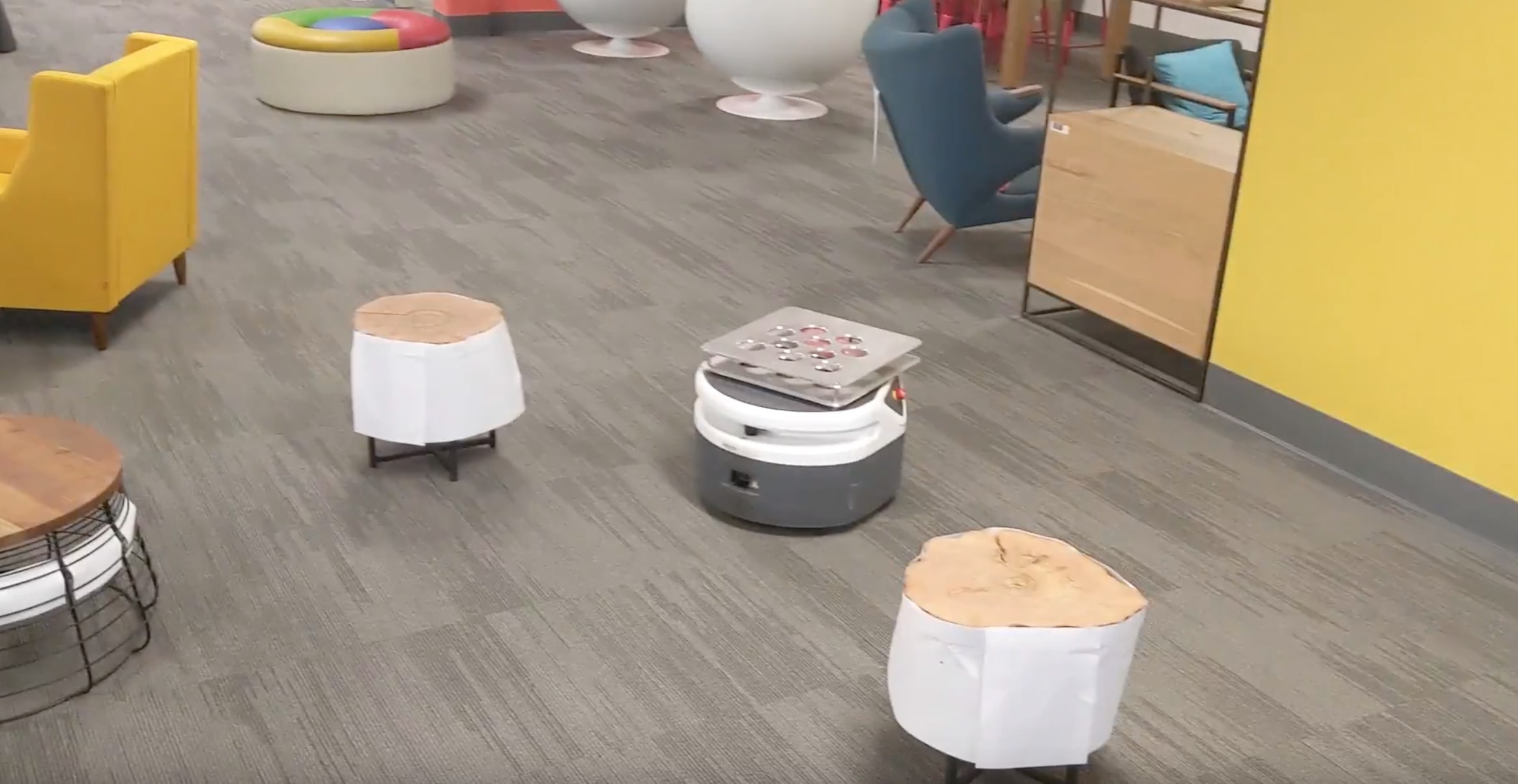}} & \parbox[c]{0.18\linewidth}{
      \includegraphics[width=1in,height=.6in]{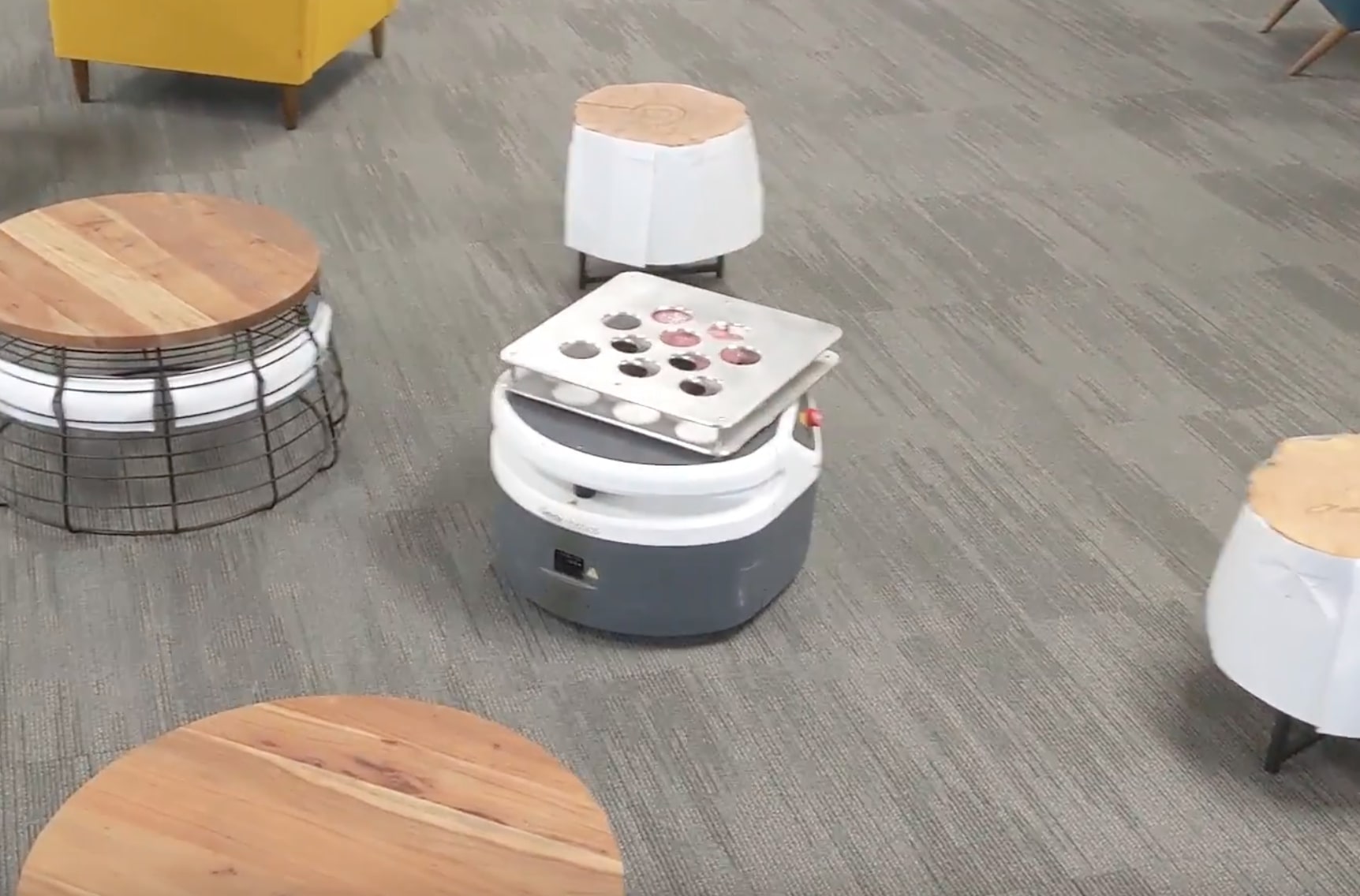}} & \parbox[c]{0.18\linewidth}{
      \includegraphics[width=1in,height=.6in]{images/real-robots2/forest3.jpg}} & \parbox[c]{0.18\linewidth}{
      \includegraphics[width=1in,height=.6in]{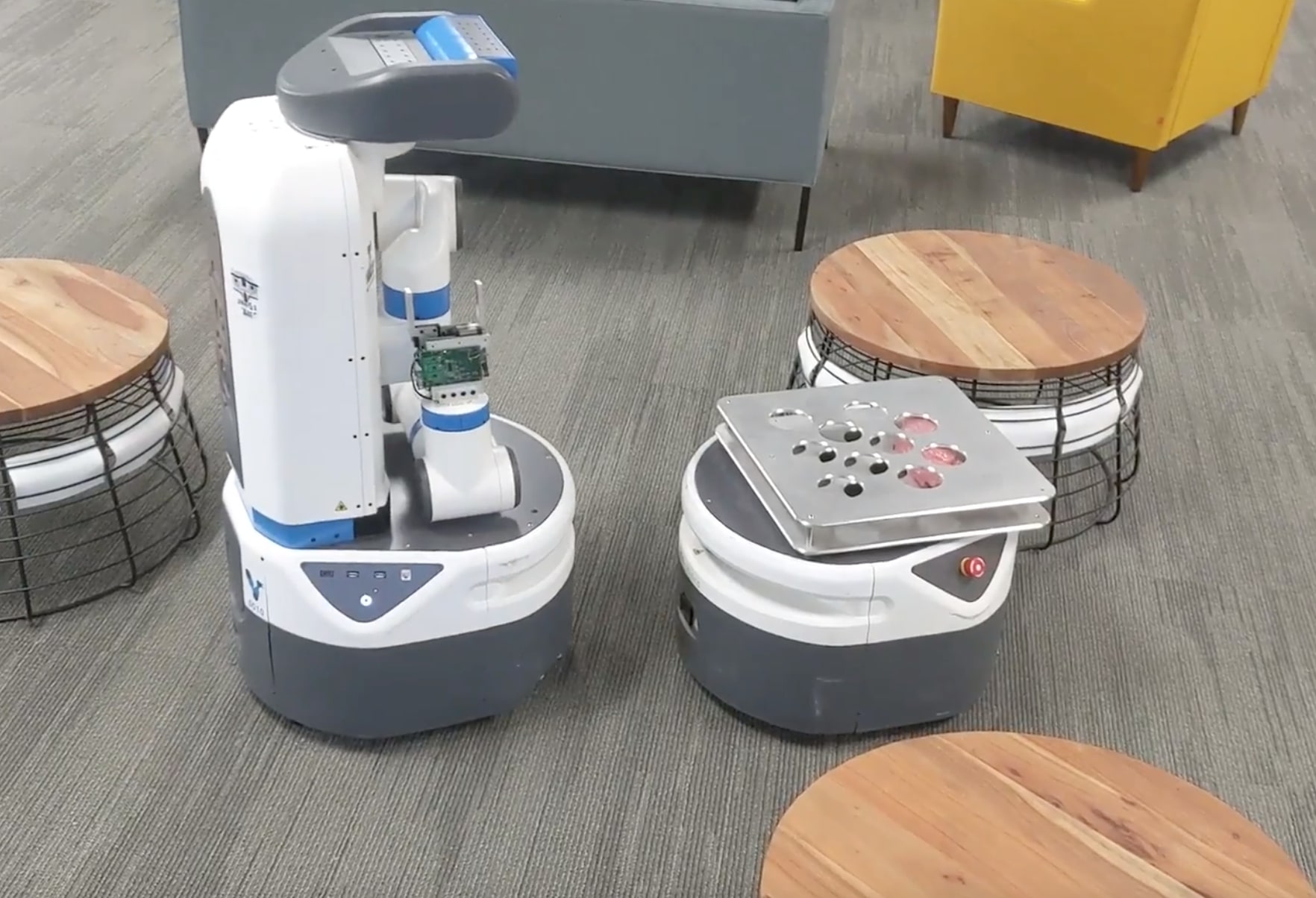}} \\
      \hline \textbf{Roundabout} & \parbox[c]{0.18\linewidth}{
      \includegraphics[width=1in,height=.6in]{images/real-robots2/round1.jpg}} & \parbox[c]{0.18\linewidth}{
      \includegraphics[width=1in,height=.6in]{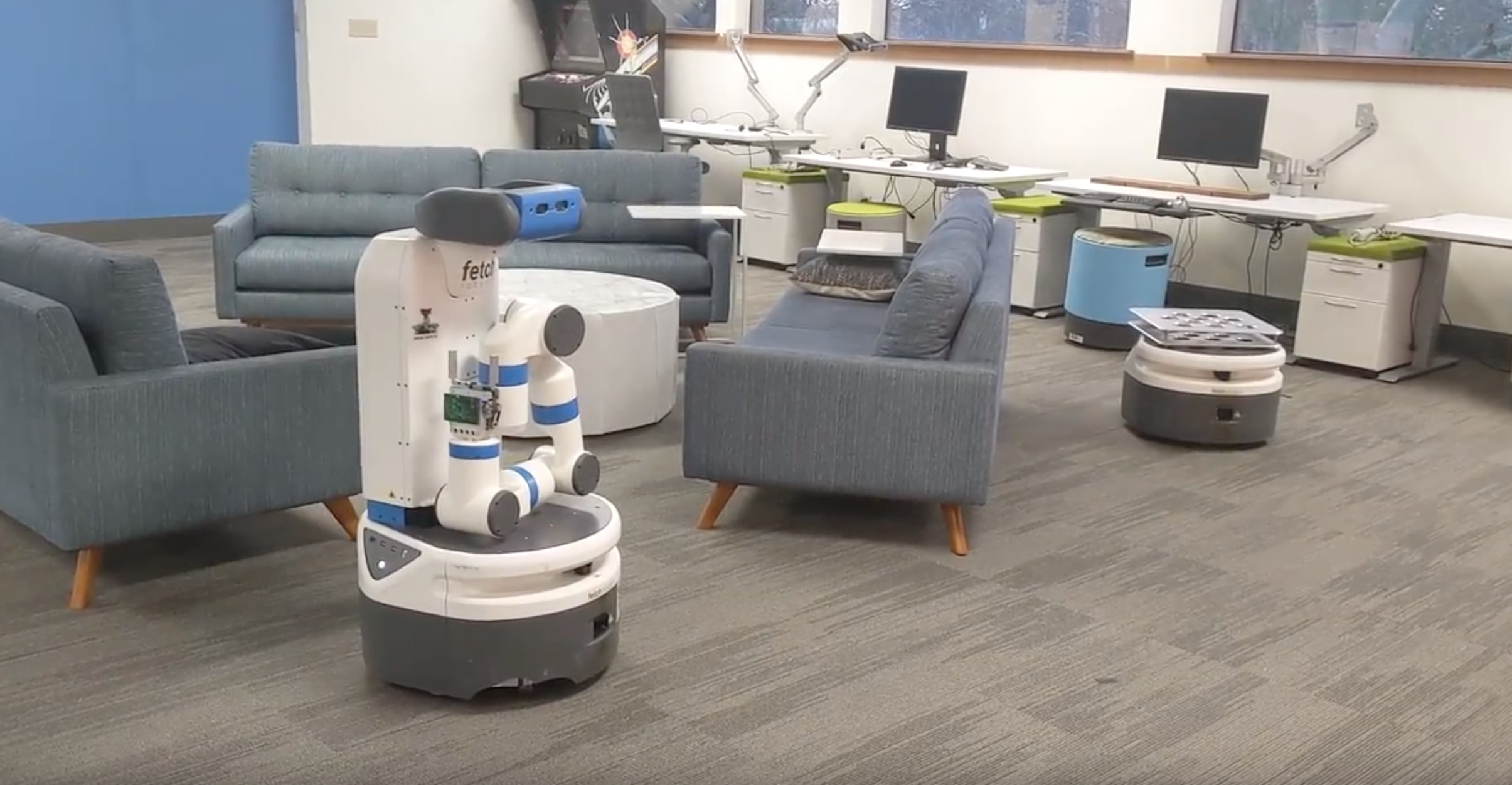}} & \parbox[c]{0.18\linewidth}{
      \includegraphics[width=1in,height=.6in]{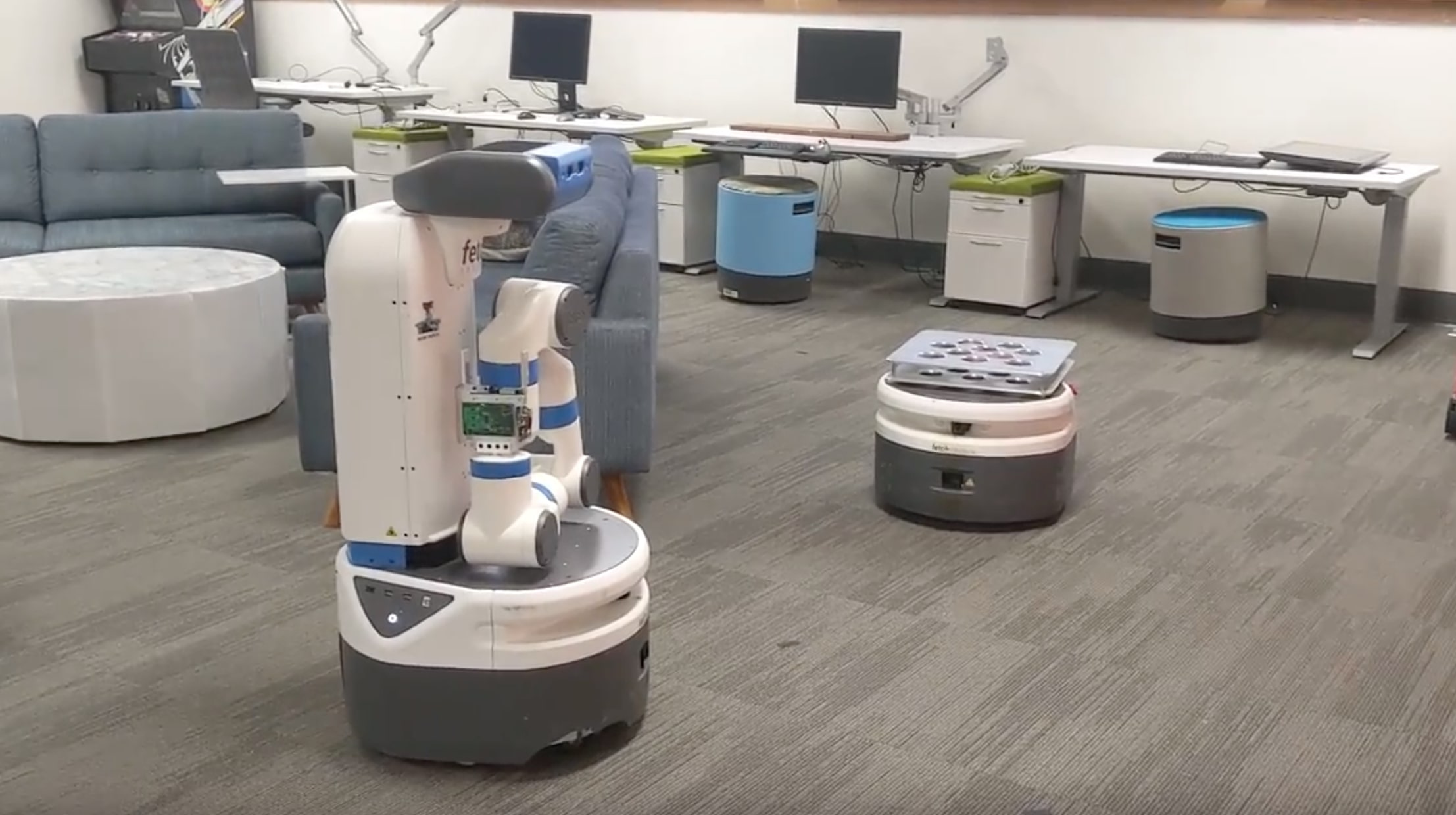}} & \parbox[c]{0.18\linewidth}{
      \includegraphics[width=1in,height=.6in]{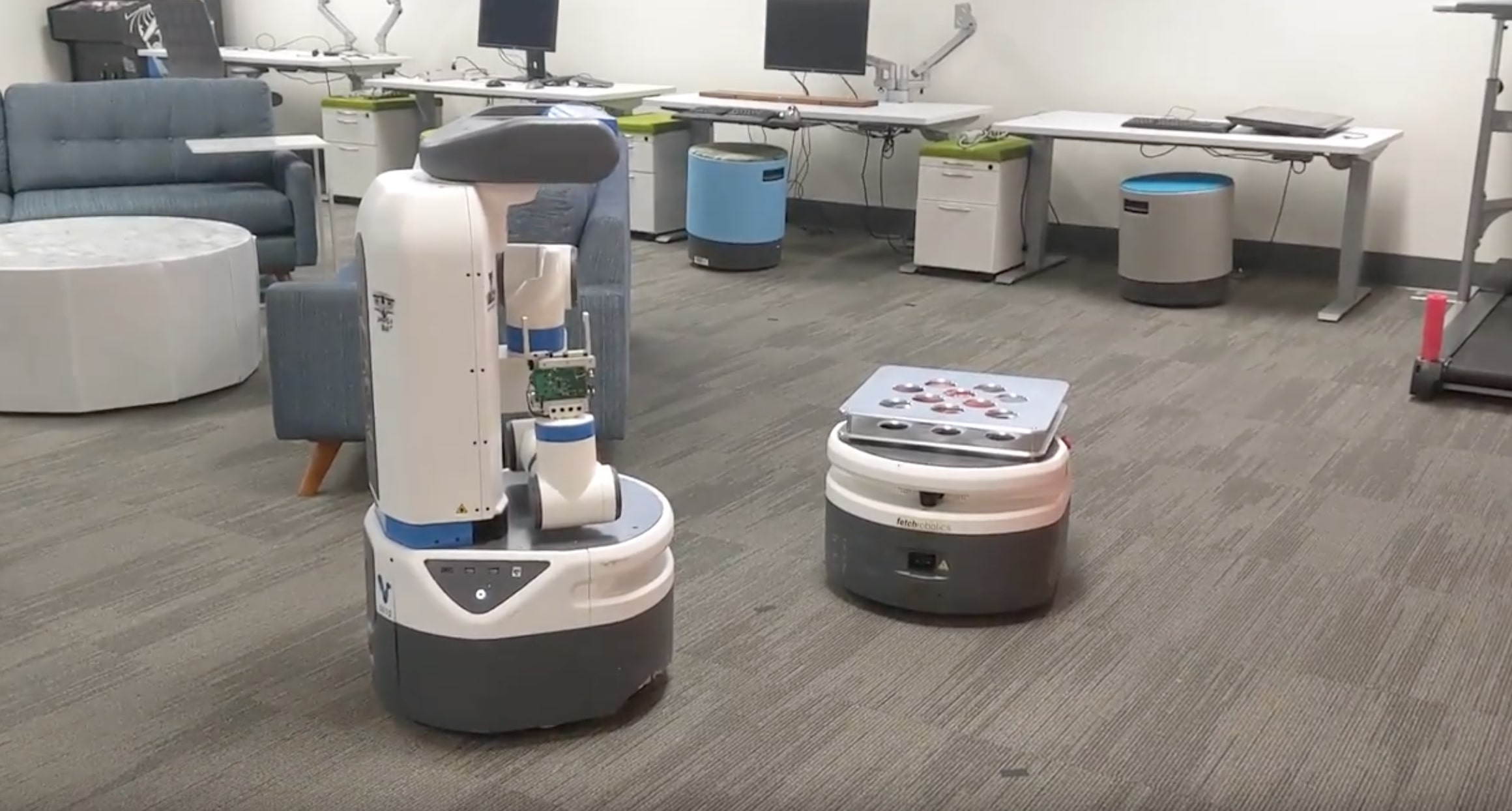}} \\
      \hline
  \end{tabular}
  \caption{Images from real world evaluations of \alg\ on different environment configurations.} \label{tab:real-world-images}
\end{table*}

\clearpage



\end{document}